\documentclass[prb,preprint,amsmath,amssymb]{revtex4}
\begin{document}
\author{Kevin Leung$^*$ and Andrew Leenheer}
\affiliation{Sandia National Laboratories, MS 1415, Albuquerque, NM 87185\\
\tt kleung@sandia.gov (505)8441588}
\date{\today}
\title{How Voltage Drops are Manifested by Lithium Ion Configurations at
Interfaces and in Thin Films on Battery Electrodes}
\input epsf

\renewcommand{\thetable}{\arabic{table}}

\begin{abstract}

Battery electrode surfaces are generally coated with electronically
insulating solid films of thickness 1-50~nm.  Both electrons and Li$^+$ can
move at the electrode-surface film interface in response to the voltage,
which adds complexity to the ``electric double layer'' (EDL).  We apply
Density Functional Theory (DFT) to investigate how the applied voltage
is manifested as changes in the EDL at atomic lengthscales, including
charge separation and interfacial dipole moments.  Illustrating examples
include Li$_3$PO$_4$, Li$_2$CO$_3$, and Li$_x$Mn$_2$O$_4$ thin-films on Au(111)
surfaces under ultrahigh vacuum conditions.  Adsorbed organic solvent
molecules can strongly reduce voltages predicted in vacuum.  We propose that
manipulating surface dipoles, seldom discussed in battery studies, may be
a viable strategy to improve electrode passivation.  We also distinguish the
computed potential governing electrons, which is the actual or instantaneous
voltage, and the ``lithium cohesive energy''-based voltage governing Li
content widely reported in DFT calculations, which is a slower-responding
self-consistency criterion at interfaces.  This distinction is critical for
a comprehensive description of electrochemical activities on electrode
surfaces, including Li$^+$ insertion dynamics, parasitic electrolyte
decomposition, and electrodeposition at overpotentials.

\vspace*{0.5in}
\noindent keywords: lithium ion batteries; voltage prediction;
density functional theory; computational electrochemistry

\end{abstract}

\maketitle

\section{Introduction} \label{intro}

\begin{figure}
\centerline{\hbox{\epsfxsize=4.00in \epsfbox{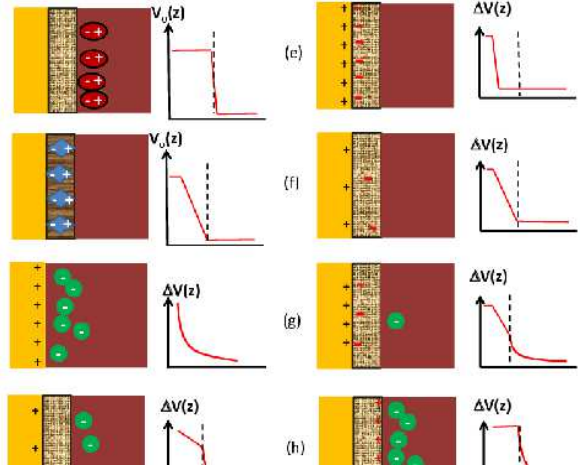} }}
\caption[]
{\label{fig1} \noindent
Some scenarios for electric double layers (EDL).  (a) Voltage drops at
electrode/liquid electrolyte interface when the metallic electrode is uncharged.
Organic solvent molecules orient themselves to reduce the vacuum voltage
(Sec.~\ref{li2co3_ec}).  (b) Solid films with fixed dipoles, like
ferroelectrics, can serve the same purpose as molecules, and represent
a novel strategy for electrode passivation (Sec.~\ref{strategy}).  (c)-(h) are
{\it additions} to intrinsic dipole-induced voltage drops of (a) when charges
exist on the cathode active material and/or thin films.  (c) Pristine, noble
metal electrode in contact with liquid electrolyte.  (d) Inert thin film
intervening between metal and liquid.  (e) EDL entirely at metallic
electrode/thin film interface (Sec.~\ref{li3po4},~\ref{li2co3}).  (f) Same as
(e), with more charge separation (Sec.~\ref{li3po4}).  (g) EDL is only
partially inside thin film (Sec.~\ref{li2co3_ec}).  (h) EDL is between
redox-active film and liquid electrolyte (Sec.~\ref{mno}).  Yellow and red
backgrounds depict Au and liquid electrolyte; light brown textures are solid
electrolyte thin films or redox-active cathode materials; green circles are
negatively charged counter ions in the liquid electrolyte.  For simplicity,
cations in the liquid are omitted.  Voltage profiles and surface charge
densities are caricatures; in the calculations, voltage changes are 1-3~V and
film thicknesses (demarcated by vertical dashed lines) are about 1~nm.  The
plots represent static conditions; ohmic losses add slopes to all flat portions.
}
\end{figure}

Unlike pristine noble metal or graphite basal-plane electrodes used in classical
electric double layer (EDL) studies,\cite{text,double} lithium ion battery
(LIB) electrodes generally exhibit complex interfaces.\cite{book3,gross1} Both
electron ($e^-$) and Li$^+$ transport can occur inside LIB electrodes.  In
addition, solid thin films, on the order 1-50~nm thick, are ubiquitous on LIB
electrode surfaces and can become part of the EDL (Fig.~\ref{fig1}a-h).  The EDL
is critical because key battery processes like Li$^+$ incorporation kinetics
(including Li$^+$ desolvation), parasitic reactions, Li-plating, and
degradation-inducing phase transformations on electrode surfaces most likely
initiate within it.  Indeed, ``solid electrolyte interphase'' (SEI)
films\cite{book3,xu,verma,korth} covering anodes are relied upon as kinetic
barriers that prevent thermodynamic equilibrium in batteries; they are critical
for LIB which operate outside the redox stability window of the organic solvent
electrolytes used.  As discussed below, the EDL is also crucial in
computational work because it effectively determines the electrode
potential (henceforth ``applied voltage'' or simply ``voltage'') in the
simulation cell.  Our work examines EDL associated with thin-film coated
electrode surfaces, and emphasizes the importance of creating electrode
interface models that exhibit consistent electrochemical activities for both
Li$^+$ and $e^-$.

Examples of solid films on electrode surfaces include Li$_2$CO$_3$ 
layers formed on pristine cathode oxide
surfaces;\cite{aurbach2000,saito2011,song2004,ostrovskii}
cathode-coating films made of electrolyte decomposition
products;\cite{book3,eriksson2002,eriksson2002a,kostecki,novak,lmo1,lmo2}
SEI films on anodes arising from reductive decomposition of liquid electrolyte
components;\cite{book3,xu,verma,korth}  
artificial protective/passivating coatings,\cite{dudney} including atomic layer
deposition (ALD) layers\cite{ald,ald1} which can undergo phase transformations
at low voltages;\cite{alotrans,ald_anode} ALD layers between solid
electrolytes and electrodes in all-solid state batteries;\cite{aldsolid}  
and even Li$_2$O$_2$ films deposited on cathodes during fast discharge
of Li-air batteries,\cite{liair,liair1} the re-oxidation of which is
accompanied by significant voltage hysteresis and is a root cause of the lack
of a suitable liquid electrolyte in Li-air batteries.  

Detailed atomic lengthscale understanding of the interfaces and EDL associated
with such thin films has remained elusive, partly due to challenges in imaging
buried interfaces at sub-nanometer resolution.  Even EDLs associated with
liquid organic electrolyte in batteries have arguably only recently received
systematic experimental studies.\cite{harris1,browning,leenheer}  Modeling
efforts have made much progress in elucidating the structures of
solid-solid\cite{sodeyama,holzwarth2,santosh}
and solid-vacuum\cite{meng,greeley,persson,ceder,islam} interfaces.  However,
voltage dependences, particularly in processes involving $e^-$ transfer like
parasitic reactions and undesirable metal plating, have arguably received
less theoretical attention.  In this work, we apply electronic Density
Functional Theory (DFT) calculations to investigate how voltages affect the
structures and stoichiometries of thin solid films, their interfaces, and
their EDL at atomic lengthscales.  In view of the complexities of LIB
electrodes, we have adopted simple model systems.  The liquid electrolyte is
omitted, although a few solvent molecules are included as frozen monolayers in
some models to illustrate their huge impact on the voltage.  Au(111) surfaces
are adopted as model cathodes, instead of transition metal oxides typically
found in batteries.  Au does not alloy with Li under voltages considered
in this work and is a convenient inert electrode.  These systems might be
realized in ultra-high vacuum (UHV) settings; they dovetail with the use of
copper\cite{harris} and gold\cite{bruce} electrodes in recent fundamental
battery science experimental studies.

\begin{figure}
\centerline{\hbox{ \epsfxsize=1.30in \epsfbox{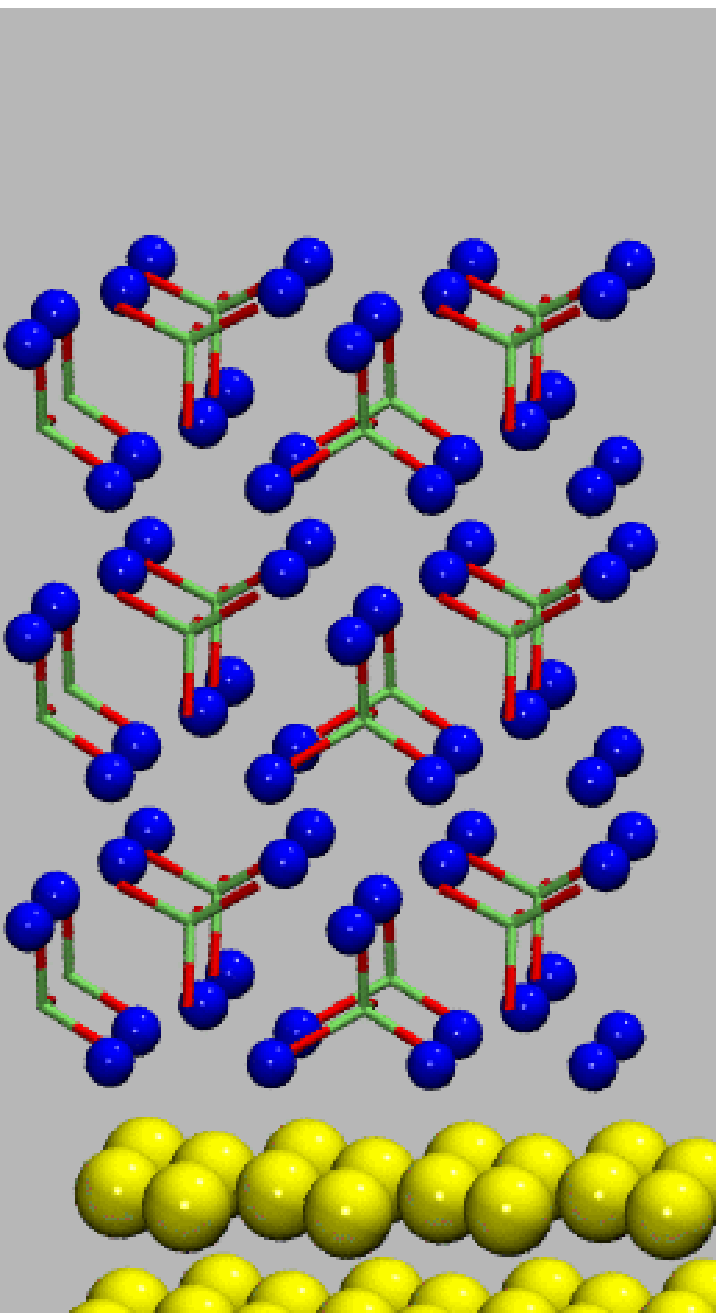} }
            \hbox{ \epsfxsize=2.40in \epsfbox{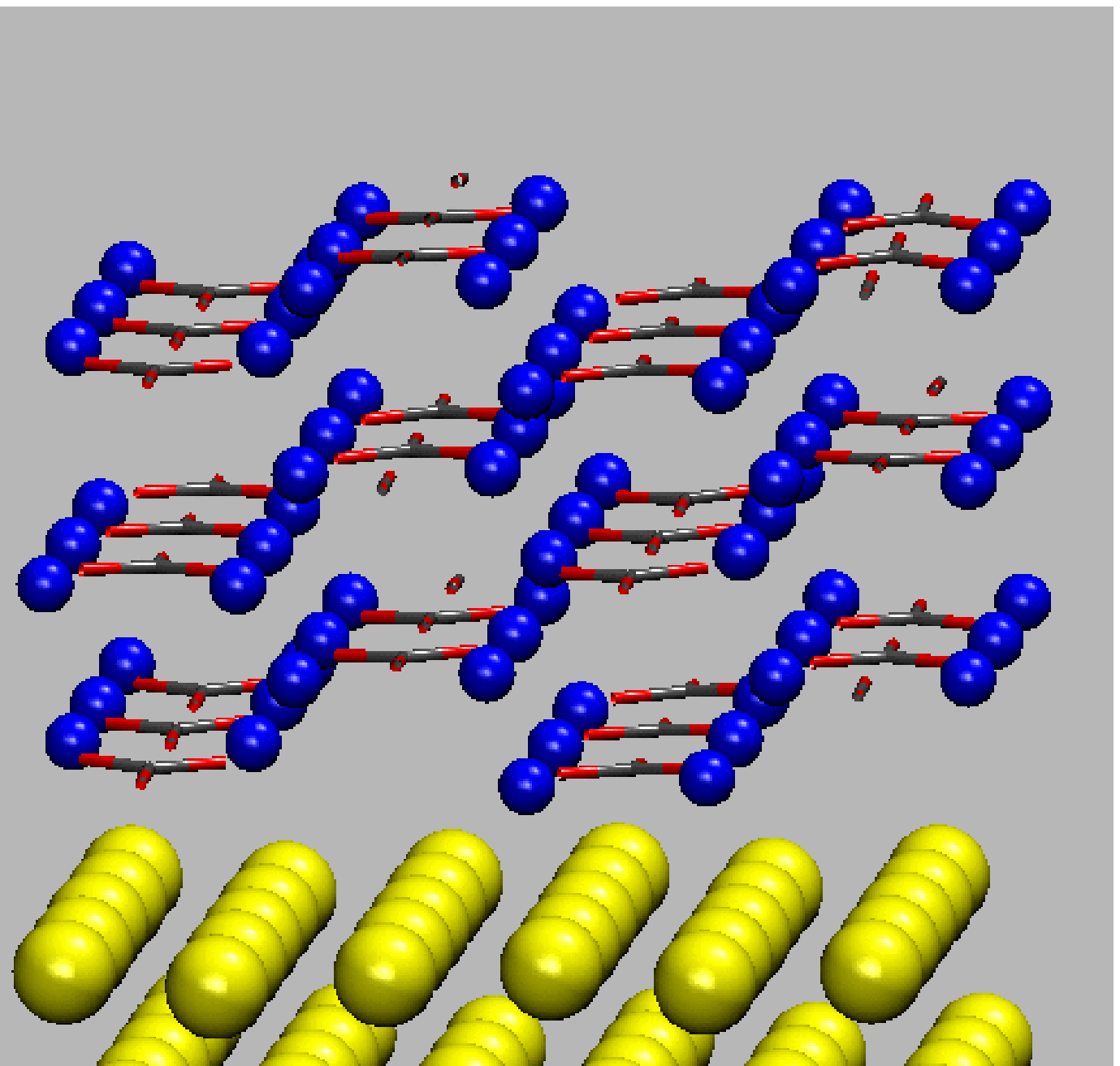} }
            \hbox{ \epsfxsize=1.28in \epsfbox{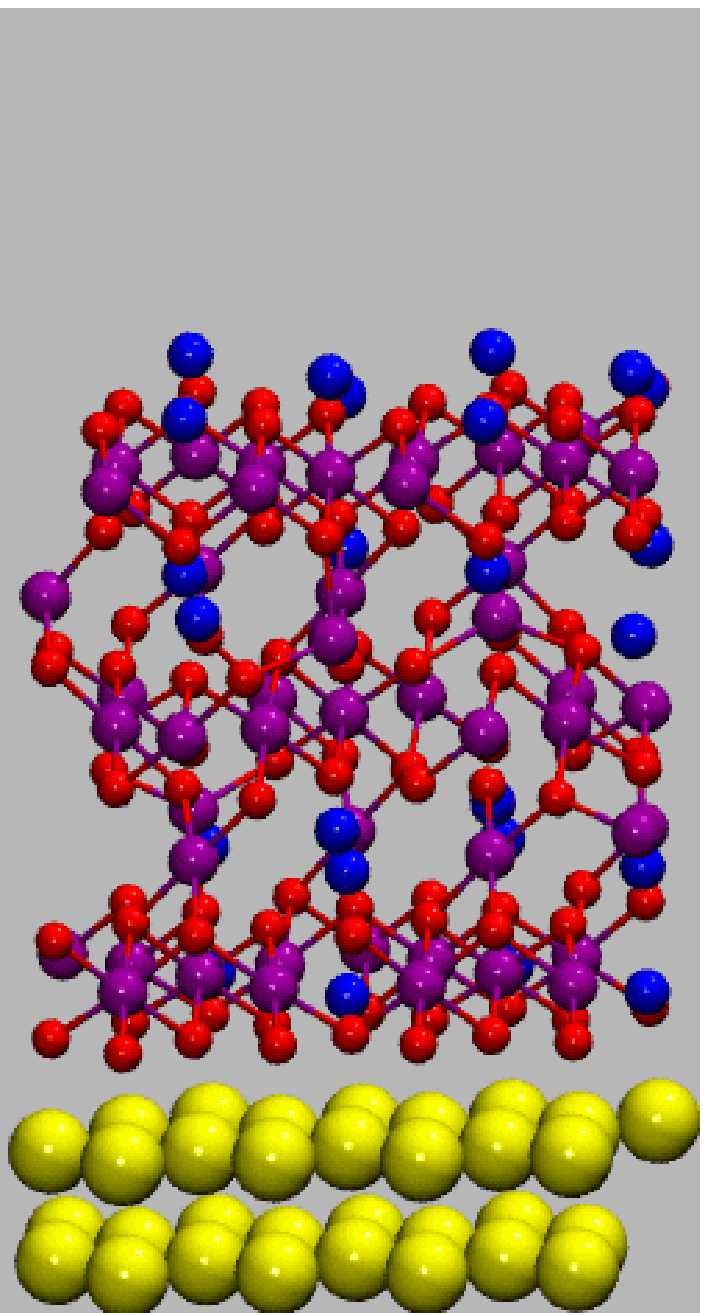} }}
\centerline{\hbox{ (a) \hspace*{0.80in} (b) \hspace*{0.6in} (c)}}
\caption[]
{\label{fig2} \noindent
(a) Li$_3$PO$_4$ (010)/Au(111).  (b) Li$_2$CO$_3$ basal plane/Au(111).  
(c) Li$_{1-x}$Mn$_2$O$_4$(111)/Au(111).  Li, P, O, C, H, Mn, and Au atoms are
depicted in blue, lime-green, red, grey, white, purple, and yellow,
respectively.
}
\end{figure}

The model thin films examined in this work, in order of increasing complexity,
are Li$_3$PO$_4$ (010) (Fig.~\ref{fig2}a),\cite{holzwarth1,holzwarth2}
Li$_2$CO$_3$ basal plane 
(Fig.~\ref{fig2}b),\cite{li2co3,li2co3_dftu,qi1,qi2,ouyang,curtiss,li2co3_surf}
and Li$_x$Mn$_2$O$_4$ (111) (Fig.~\ref{fig2}c).\cite{persson}  These are coated
on Au(111) on one side and face a vacuum region on the other.  Li$_3$PO$_4$ is
a non-redox-active solid electrolyte.  It illustrates the correlation between
interfacial dipole densities and voltages.  Li$_2$CO$_3$ is often found on
as-synthesized cathode surfaces.  In LIB studies, there is
disagreement\cite{saito2011} concerning whether Li$_2$CO$_3$ dissolves upon
soaking in electrolyte,\cite{aurbach2000,song2004} is formed at high
voltages,\cite{ostrovskii} or is removed at $>$4~V.\cite{saito2011}  In
Li-air batteries, Li$_2$CO$_3$ is generally accepted to be oxidized and
removed above $\sim$4.5~V.\cite{liair1}  Our goal is not to elucidate the
detailed reaction mechanism, but to study the electronic and structural
signatures accompanying voltage increase.  Redox-active LiMn$_2$O$_4$ (spinel
LMO) is not used as nanometer-thick films in batteries, but its inclusion here
helps illustrate the diversity of interfacial behavior (Fig.~\ref{fig1}).
Like most LIB cathode materials, LMO is a polaron conductor because Mn can
exhibit multiple redox states.  It also conducts Li$^+$, but is not a band
(``metallic'') conductor.  By classical electrostatics definition, the interior
of LMO is not instantaneously at a constant potential.\cite{kanno}  Our LMO/Au
interface model provides a well-defined Fermi level ($E_{\rm F}$) to
demonstrate how voltage control is achieved at atomic lengthscales, and how
this affects EC oxidative decomposition on LMO surfaces.\cite{lmo2}

Out of necessity, we critically examine the voltage calibration method used
in the theortical LIB literature (Sec.~\ref{voltage}).  The same experimental
voltage governs Li$^+$ and $e^-$ motion.  Indeed, in cyclic voltametry, it is
difficult to distinguish currents arising from Faradaic Li-insertion or
parasitic processes that involve undesired $e^-$ transfer to the liquid.  
But what is ``voltage'' at atomic lengthscales?  The absolute electrochemical
potential for a charged particle in an atomic configuration is the (free)
energy needed to transfer it into a given phase from infinity (see, e.g.,
Ref.~\onlinecite{rossmeisl2015}).  There can be differences
whether the particle is a Li$^+$ or a $e^-$.  In DFT modeling, the electronic
voltage (${\cal V}_e$) governing electronic motion, and ionic voltage
(${\cal V}_e$) governing Li$^+$ motion, can and must be distinguished.

First we focus on ${\cal V}_e$.  Under constant voltage experimental conditions,
potentiostats directly control the $e^-$ (not Li) content.  When the phase is a
metallic conductor with a well-defined $E_{\rm F}$, the voltage asociated with
transferring $e^-$ from infinity, multiplied by the electronic charge, is equal
to the work function ($\Phi$) modulo
a constant, just like in photovoltaics (PV)\cite{brocks}  and
electrocatalytic systems.\cite{gross} $\Phi$=$E_{\rm vac}-E_{\rm F}$, where
$E_{\rm vac}$ is the vacuum level and $E_{\rm F}$ is set to zero herein without
loss of generality.  If the electrode is immersed in liquid electrolyte, $\Phi$
calculated for the electrode covered by a sufficiently thick liquid layer,
divied by the electronic charge, also gives ${\cal V}_e$.\cite{sprik,otani08}
${\cal V}_e$ applies
even under out-of-equilibrium conditions.  It is readily computed in our model
systems because vacuum regions exist and qualify as ``infinity.'' In commercial
batteries, there is no vacuum.  Nevertheless, our focus on $\Phi$ is formally
correct and emphasizes the synergy between batteries, fuel
cells,\cite{rossmeisl15} PV,\cite{pv1,pv,nsai,brocks} and metal-semiconductor
interfaces.\cite{tung1,tung2}  In those research areas, dipoles at interfaces
are recognized as properties of paramount importance.  At battery liquid-solid
interfaces, EDL's are the manifestations of complex surface dipole distributions
associated with applied voltages; it is in effect our computational
potentiostat.  Based on our calculations, we propose the manipulation of
surface dipoles as a novel electrode passivating strategy (Fig.~\ref{fig1}b).

Li$^+$ ions are only indirectly controlled by potentiostats; they slowly
redistribute in response to changes in electronic configurations.  We define
the Li ``voltage'' (${\cal V}_i$) as the (free) energy difference between Li
chemical potential in a system and the lithium metal cohesive energy per Li
divided by $|e|$.  This definition is widely used as the sole voltage estimate
in the LIB theory literature.\cite{licohes1,licohes2,ceder_review,meng_review}
At equilibrium, when the voltage is pinned at the redox potential of
Li-insertion reactions, ${\cal V}_e$ must be equal to ${\cal V}_i$.  Thus
${\cal V}_i$ should be considered a self-consistent criterion at interfaces.
By itself, ${\cal V}_i$ can give incorrect voltages when modeling parasitic
processes induced by $e^-$ transfer, and it may not govern the dynamics of
Li$^+$ insertion processes adequately.  

\begin{figure}
\centerline{\hbox{ \epsfxsize=3.50in \epsfbox{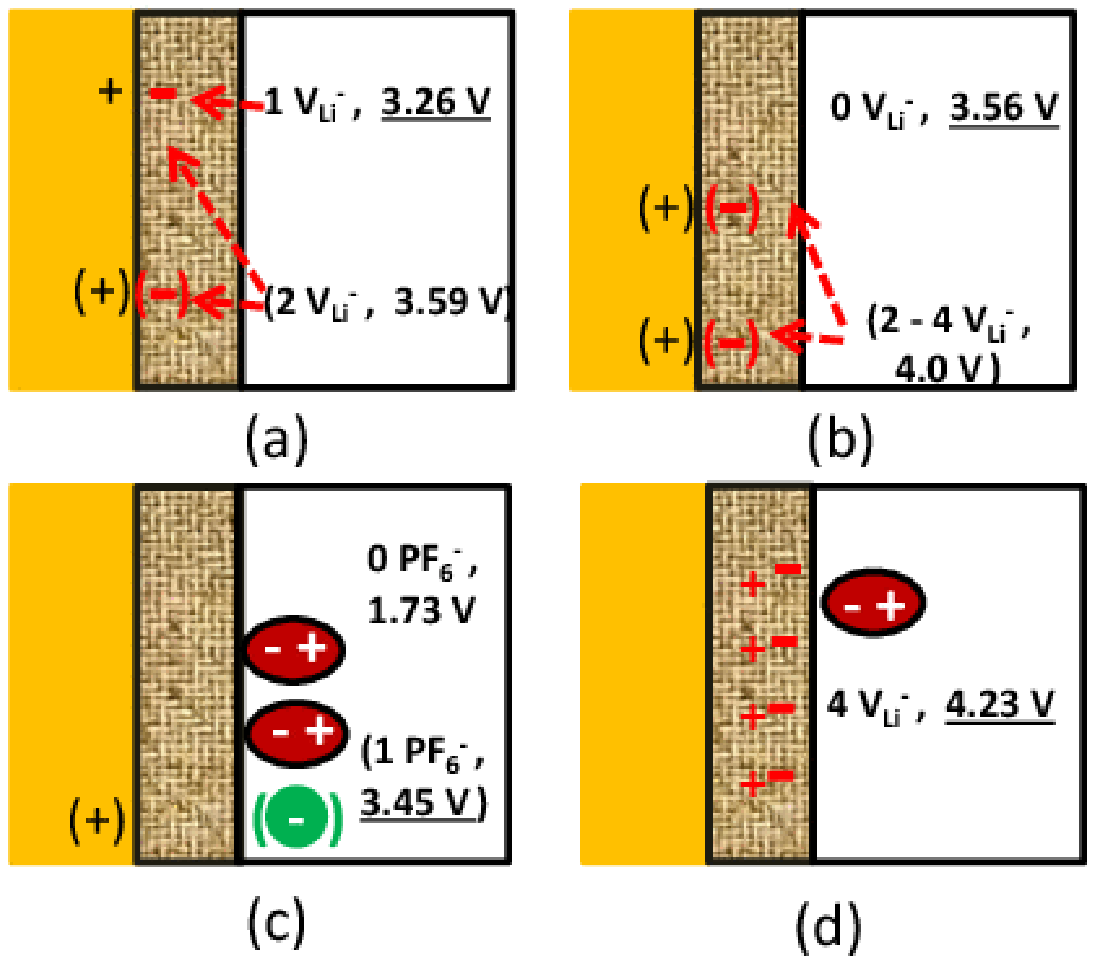} }}
\caption[]
{\label{fig3} \noindent
Overview of representive predictions of how EDL affects electronic
voltage (${\cal V}_e$).  The white background is vacuum.  Yellow and textured
regions are Au and solid thin films, respectively:
(a) Li$_3$PO$_4$/Au (Sec.~\ref{li3po4});  
(b) Li$_2$CO$_3$/Au (Sec.~\ref{li2co3});  
(c) EC+PF$_6^-$/Li$_2$CO$_3$/Au (Sec.~\ref{li2co3_ec});  
(d) EC/Li$_x$Mn$_2$O$_4$/Au (Sec.~\ref{mno}).
``V$_{\rm Li}^-$'' are negatively charged Li$^+$ vacancies.  ${\cal V}_e$
values with and without parentheses are two alternate scenarios.
Underlined values are at equilibrium (i.e., ${\cal V}_e$=${\cal V}_i$).
PF$_6^-$ anions on Li$_x$Mn$_2$O$_4$ surfaces are found to accelerate
EC decomposition in their vicinity (not shown; Sec.~\ref{ec_decomp}).
}
\end{figure}

Fig.~\ref{fig3}a-d offers a preview of our results, pertinent to issues
discussed above.  The rest of the paper is organized as follows.  Sec.~2
discusses and reconciles different definitions of voltage in DFT calculations
in more detail.  Sec.~3 describes the methods used.  The results are given
in Sec.~4 and Sec.~5 summarizes the paper. A supporting information (S.I.)
document provides further details about our three model systems, compares the
modeling of battery interfaces with
geochemical\cite{geochem,geochem1,multiscale} and other liquid-oxide
interfaces,\cite{adriaanse,sodeyama1} and discusses Li$_2$CO$_3$ oxidation
thermodynamics.  Experimentalist readers are encouraged to skip to
Sec.~\ref{results}.

\section{Voltage Calibration in DFT Calculations} \label{voltage}

DFT calculations are conducted at constant number of electrons, not constant
electrochemical potential.  Estimating and maintaining voltages in
periodically replicated, condensed phase DFT simulation cells have long been
recognized as a challenge.\cite{sprik,otani12,otani13,rossmeisl13,rossmeisl15,gross,arias,neurock,norskov}
Recently we have applied an {\it ab initio} molecular dynamics (AIMD)-based
Li$^+$ transfer free energy method to predict voltage dependences of redox
processes at interfaces between liquid electrolytes and lithium-intercalated
graphite.\cite{voltage,vedge}  AIMD has also been coupled with continuum
approximations of liquid electrolytes to model constant voltage conditions
on pristine electrodes.\cite{otani12}  But present-day AIMD simulation time
scales are too short to extend to Li$^+$ motion {\it inside} LIB electrodes or
the surface films coating them except at significantly elevated temperatures.
This work focuses on how applied voltages affect solid films.

Thus we model solid interfaces at zero temperature.  With sufficient
equilibration time and in the presence of Li$^+$ and $e^-$ reservoirs,
the total number of electrons ($n_e$) and Li$^+$ content ($n_{\rm Li+}$)
in an electrode in a LIB are governed by the chemical potentials for electrons
($\mu_e$) and Li$^+$ ions ($\mu_{\rm Li+}$), respectively.  $E_{\rm F}$ of a
metallic electrode is $\mu_e$ modulo a constant.  Here we have not used 
${\bar \mu}_e$ or ${\bar \mu}_{\rm Li+}$, the notations for total
electrochemical potentials of $e^-$ and Li$^+$,\cite{text} precisely
because constant-charge DFT calculations and terminologies are being applied
to infer constant-potential electrochemical properties.  Surface potential
contributions to ${\bar \mu}$'s should naturally be present
in our simulation cells, which feature explicit interfaces.  

\subsection{Electronic Voltage (${\cal V}_e$) from Work Function}

The electronic voltage of an atomic configuration which is band conductor,
imposed by a potentiostat, not necessarily at equilibrium with respect to
$n_{\rm Li}$ or the Li$^+$ occupation sites, is defined as
\begin{equation}
{\cal V}_e=\Phi/|e|-1.37~{\rm V}, \label{instant}  
\end{equation}
if the Li$^+$/Li(s) reference is used.\cite{borodin137}  In the literature,
values of 1.39~V and 1.44~V have been adopted; the small discrepancy is a
measure of possible systematic error.  In the DFT formulation, the $e^-$
configuration finds its ground state instantaneously.  ${\cal V}_e$ pertains
to infinitesimal changes in $n_e$, which leaves the system uncharged.  Note
that Refs.~\onlinecite{voltage} and~\onlinecite{vedge} deal with ${\cal V}_e$
despite the fact that Li$^+$ is transferred, because the excess $e^-$ is left
on the metallic electrode at its Fermi level.  As stated in
Ref.~\onlinecite{vedge}, the free energy of that graphite edge-plane system
has yet to be optimized with respect to the surface Li-content.  In the present
work, ${\cal V}_e$ are always reported for locally optimized atomic
configurations so that all forces are zero.  

\subsection{Lithium Metal Cohesive Energy (${\cal V}_i$) for Voltage}

In the ``lithium cohesive energy'' (${\cal V}_i$) approach, discrete
and matched numbers of $e^-$ and Li$^+$ are added/subtracted simultaneously:
\begin{eqnarray}
\mu_{\rm Li} &=& E(n_{\rm Li+},n_e)-E(n_{\rm Li+}-1,n_e-1) \nonumber \\
	&=& E(n_{\rm Li+},n_e)-E(n_{\rm Li+}-1,n_e) +
	 E(n_{\rm Li+},n_e)-E(n_{\rm Li+},n_e-1) \nonumber \\
	&\approx& \mu_{\rm Li+} + \mu_e . \label{mu_li}
\end{eqnarray}
The last equality holds when the system approaches infinite size.  Here
$E(n_{\rm Li+},n_e)$ is the zero temperature total energy after optimization
of all interior degrees of freedom.  The $T=0$~K voltage is then given by 
\begin{equation}
{\cal V}_i = (\mu_{\rm Li}-E_{\rm Li(s)})/|e|, \label{relax}
\end{equation}
where $E_{\rm Li(s)}$ is the cohesive energy of lithium metal, predicted to
be 1.57~eV using the DFT/PBE functional.  No net charge is introduced.  There
is an ambiguity concerning whether $\mu_{\rm Li}(n_{\rm Li})$ should be 
$E(n_{\rm Li}+1)-E(n_{\rm Li})$ or $E(n_{\rm Li}1)-E(n_{\rm Li}-1)$.  We have
chosen the latter definition (Eq.~\ref{mu_li}).  $\mu_{\rm Li}$ is used
instead of $\mu_{\rm Li+}$ to control $n_{\rm Li}$ because the former is the
widely used convention, and because it is harder to control charged particles
due to the periodic boundary condition adopted.  In the literature,
${\cal V}_i$ is generally averaged over a wider range of Li content than
in the present work.\cite{licohes1,licohes2}

${\cal V}_e$ and ${\cal V}_i$ are not necessarily equal.  A transparent example
is a pristine Li metal slab in vacuum.  ${\cal V}_i$=0 by definition.  Yet the
work function of Li(100) is reported to be 2.93~eV.\cite{crc} This translates
into an electronic voltage of ${\cal V}_e$=1.51~V (Eq.~\ref{instant})!  Indeed,
if the Li surface is covered with a poor SEI that allows $e^-$ transfer but
Li$^+$ transfer is slowed to beyond experimental timescales -- the opposite
of what a good SEI does -- the lithium slab acts as an $e^-$ emitter, the
voltage of which must be governed by its $E_{\rm F}$.  For Li(s) to truly
exhibit 0~V vs.~the LIB Li$^+$/Li(s) reference in $e^-$ transfer processes,
net surface charge densities and electrode-electrolyte interface contributions
must be accounted for.  Organic solvent decomposition can occur on uncharged
Li (100) surface under UHV conditions despite the relatively high (1.51~V)
potential, but in these reactions the metal surface is a reactant and Li$^+$
are produced.

\subsection{Relation between ${\cal V}_e$ and ${\cal V}_i$}

(1) ${\cal V}_e$=${\cal V}_i$ at equilibrium.  Under this condition,
${\cal V}_i$ controls the Li content.  Thus Eq.~\ref{relax} should be
interpreted as a self-consistency criterion.  
$\mu_{\rm Li}$=($\Phi$-1.37~eV+1.57~eV) from Eqs.~\ref{instant},~\ref{relax},
and the PBE estimate of $E_{\rm Li}(s)$.  Identifying $\Phi$ as $\mu_e$,
Eq.~\ref{mu_li} implies that $\mu_{\rm Li+}$ becomes a voltage-independent
0.20~eV.  This reflects that Li$^+$ is in excess in an infinite reservoir.  
A related point has been made in the fuel cell literature.\cite{rossmeisl15}
In that case H$^+$ concentration at the interface, governed by the pH, is
also strongly coupled to the predicted voltage.

(2) When ${\cal V}_i$$<$${\cal V}_e$, and the voltage is held fixed
by the potentiostat, and Li$^+$ and $e^-$ sources are present, the Li content
should spontaneously decrease (as permitted by kinetics) so that ${\cal V}_i$
rises towards ${\cal V}_e$.  But ${\cal V}_i$ does not always increase with
decreasing $n_{\rm Li}$.  Going back to the lithium solid example: if the
potentiostat is set at 1.51~V in solution, the entire Li slab dissolves.  If
instead an electrically disconnected Li solid is dipped into an electrolyte,
some Li$^+$ ionizes, leaving a negatively charge, possibly SEI-covered, Li
surface.  Now it is the electronic voltage ${\cal V}_e$ which decreases to
0~V vs.~Li$^+$/Li(s) to coincide with ${\cal V}_i$.

(3) When ${\cal V}_i$$>$${\cal V}_e$ and the voltage is held fixed, the
Li content should in general increase to lower ${\cal V}_i$.  This may not
occur in all cases.  For example, when all available Li$^+$ insertion sites
are filled in fully lithiated LiC$_6$, the electrode becomes
supercapacitor-like.  ${\cal V}_e$ now controls the net
surface electronic charge,\cite{voltage} and ${\cal V}_i$ ceases to matter.
Conducting simulations under such non-equilibrium conditions are crucial to
understanding many phenomena in batteries.  An example is the kinetics of
metal plating\cite{plating,gross1,zavadil} at overpotentials -- an explicitly
non-equilibrium phenomenon, a major safety issue in LIB, but also a critical
process for batteries featuring rechargeable metal anodes.  This
discussion highlights the primacy of ${\cal V}_e$ when computing voltages.

Eq.~\ref{relax} was originally applied to bulk crystalline electrode
simulation cells, and ${\cal V}_i$ has been identified as the redox
potential.\cite{licohes1,licohes2,zhou} Li should remain in the solid when
the applied potential is lower than ${\cal V}_i$.  In these special cases of
simulation cells without interfaces, the Li(s) cohesive energy approach
(${\cal V}_i$) is {\it neither consistent nor inconsistent} with ${\cal V}_e$.
This is because the electrostatic potential at any point in space is only
defined to within a constant in the interior of a crystal;\cite{electr}
therefore neither ${\cal V}_e$ nor $E_{\rm F}$ is well-defined without
specifying interfaces.  Thus metals terminated in different crystal facets
exhibit different $\Phi$'s.\cite{marzari} Much more significant changes
in $\Phi$ can be induced using adsorbed molecules.\cite{brocks}
Using DFT+U or hybrid DFT functional methods, ${\cal V}_i$ has generally been
predicted to be in good agreement with measured open circuit voltages in
bulk transition metal oxide simulation cells.  The errors are at most a few
hundred meVs.\cite{zhou} This suggests that interfacial contributions may be
small on most battery cathode surfaces.  Our work suggests that artificial
interfaces can be potentially be engineered to give larger, beneficial effects.
Molecular adsorption effects will be emphasized in Sec.~\ref{li2co3_ec}.

DFT calculations of interfaces are usually performed in finite-sized,
periodically-replicated simulation cells kept at overall charge neutrality.
In the literature, in DFT calculations without ions in the electrolyte (or
without any electrolyte at all), the ${\cal V}_i$ approach has therefore
assigned a wide range of voltages to LIB interfaces at charge neutrality.  In
other words, a continuous range of potentials-of-zero-charge (PZC) have been
assigned by DFT.  Yet simple ``classical'' electrodes such as pristine noble
metals usually permit one PZC;\cite{text} at voltages away from PZC, the
surface is charged.  If liquid electrolyte is present, the electrode surface
can exhibit non-zero charges, compensated by counterions in the electrolyte.
In this work, we illustrate this effect by adding PF$_6^-$ anions under UHV
conditions.

\section{Methods} \label{method}

DFT calculations are conducted using the Vienna Atomic Simulation Package
(VASP) version 5.3\cite{vasp,vasp1,vasp2} and the PBE functional.\cite{pbe}
Modeling spinel Li$_x$Mn$_2$O$_4$ requires spin-polarized DFT and the
DFT+U augmented treatment\cite{dftu} of Mn $3d$ orbitals.  The $U$ and $J$
values depend on the orbital projection scheme and DFT+U implementation
details; here $U-J=$4.85~eV is chosen to taken from the literature.\cite{zhou}  
The charge state on Mn ions are determined from the approximate local
spin polarzation $s_z$ reported by the VASP code; $|s_z|$=3/2,~4/2, and~5/2 are
assigned to Mn(IV), Mn(III), and Mn(II), respectively.  For Li$_3$PO$_4$ and
Li$_2$CO$_3$, most calculations exclude spin-polarization.  A few
spin-polarized calculations are performed to confirm that this is adequate.
A 400~eV planewave energy cutoff and a 3$\times$3$\times$1 Monkhorst-Pack
grid are applied in all cases.  Increasing the cutoff or the $k$-grid changes
the predicted absolute energies by at most 0.1~eV; energy differences, which
are the relevant quantities, are much smaller.

Our model systems are asymmetric slabs: Li$_2$CO$_3$ basal plane, Li$_3$PO$_4$
(010), and LiMn$_2$O$_4$ (111) slabs on one side of 4-layer thick Au(111)
(Fig.~\ref{fig2}).  The standard dipole correction is applied to
negate image interactions in the periodically replicated, charge-neutral
simulation cells.\cite{dipole_corr} Au is a metallic conductor and its
$E_{\rm F}$ is well defined.  Work functions ($\Phi$) are the differences
between $E_{\rm F}$ and vacuum levels.  Two vacuum-surface interfaces exist
per asymetric slab, and two vacuum levels, for bare and coated Au(111)
respectively, are obtained.\cite{brocks}  Four layers of Au atoms do not give
a completely converged $\Phi$ for uncoated Au(111).  Therefore we have shifted
the two computed $\Phi$ of each system by the difference between the predicted
Au(111) $\Phi$, and the fully converged Au(111) $\Phi$=5.15~eV.\cite{au_note}
The shift is at most 0.1~eV.  We also report the net surface dipole density
($\delta$), which is the total dipole moment of the simulation
cell in the direction perpendicular to the surface divided by the lateral
surface area.  The dipole moment is computed as the sum of all charges
(including electron density defined on a grid) multipled by their displacement
from a user-specified center of the cell, and is reported by the VASP code.
$\delta$ should be independent of the cell-center in a charge-neutral
simulation cell as long as the choice locates the artificial dipole
layer\cite{dipole_corr} inside the vacuum region.   More details on the thin
film models, and rationale for crystal facets chosen, are provided in the S.I.

\section{Results} \label{results}

\subsection{Au(111)/Li$_3$PO$_4$/Vacuum (Fig.~\ref{fig3}a for Summary)}
	\label{li3po4}

\begin{table}\centering
\begin{tabular}{ |c|r|r|r|r|r|r|r|  } \hline
$\Delta N$(Li) & 0 & -1 & -2 & -3 & -4 & -1$^a$ & -1$^b$ \\ \hline
${\cal V}_e$ & 2.86 & 3.26 & 3.59 & 3.92 & 4.19 & 3.76 & 4.59 \\
${\cal V}_i$ & 3.07 & 3.23 & 3.58 & 3.65 & NA & (3.69) & (4.10) \\ \hline
\end{tabular}
\caption[]
{\label{table1} \noindent
Electronic (${\cal V}_e$) and ionic (${\cal V}_i$) voltages of a
Li$_3$PO$_4$(010)/Au(111) interface cell as the number of Li vacancies at the
the inner surface of the Li$_3$PO$_4$ slab varies.  Exceptions:
cases $a$ and $b$ involve removing a Li from a middle layer or from the
outer surface of Li$_3$PO$_4$.  ${\cal V}_i$ is defined for incremental Li 
subtraction, e.g., the ``-1'' value refers to the difference between the
total energies of ``-1'' and ``-2.''
}
\end{table}

\begin{figure}
\centerline{\hbox{ \epsfxsize=3.50in \epsfbox{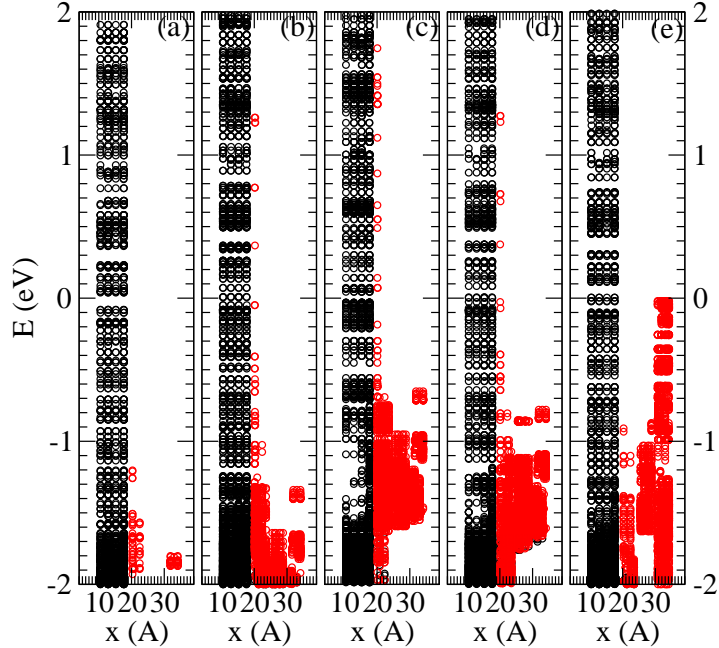} }}
\caption[]
{\label{fig4} \noindent
Kohn-Sham orbitals of Li$_{3-x}$PO$_4$/Au decomposed on to atoms at
their coordinates perpendicular to the interface.  $\Delta N$(Li) is
(a) 0; (b) $-1$; (c) $-3$;
(d) $-1$$^a$; (e) $-1$$^b$.  The vacancies are in the phosphate layer closest
to Au(111) (Fig.~\ref{fig5}a-b) except for panel (d) (case $a$, middle of
slab, Fig.~\ref{fig5}c) and (e) (case $b$, vacancy in outermost layer, not
depicted in Fig.~\ref{fig5}).  $E_{\rm F}$ is shifted to zero in each case;
vacuum levels are not aligned, and are at 4.23, 4.63, 5.29, 5.16, and 6.19~eV
in the five panels.  Au and thin-film-based orbitals are depicted in black
and red, respectively.  $\rho_c$=0.007 is used for determining orbital
locations.\cite{note7}
}
\end{figure}

First we examine the ${\cal V}_e$ of a non-redox active system: a thin
$\beta$-Li$_3$PO$_4$ (010) solid electrolyte film on Au(111).  We start with
the stochiometric Li$_3$PO$_4$ slab with all Li$^+$ sites occupied
($\Delta N$(Li)=0, Fig.~\ref{fig2}a).  The work function is $\Phi$=4.23~eV.
This translates into ${\cal V}_e$=2.86~V vs.~Li$^+$/Li(s), which is reduced
from the bare, charge-neutral Au(111) surface value ${\cal V}_e$=3.78~V
($\Phi$=5.15~V) by 0.88~eV.  In the absence of Au, the Li$_3$PO$_4$ slab is
symmetric and does not exhibit a net dipole moment.  The fact that it modifies
the Au $\Phi$ is not due to charge transfer between Au and Li$_3$PO$_4$.
The spatially decomposed electronic density of state (DOS) (Fig.~\ref{fig4}a)
shows that the phosphate Kohn-Sham valence band edge, located at about
$-1.5$~eV of the phosphate,\cite{note7} is far from the Fermi level residing
on Au orbitals.  The Li$_3$PO$_4$ conduction band edge is higher than 5~eV and
the vacuum level is at 4.23~eV (not shown in the panel).  

Thus the stoichiometric slab must have induced image charges which set up
surface dipoles.  Indeed, the net surface dipole density ($\delta$) of the
entire simulation cell is predicted to be finite; $\delta$=$-$0.0048~$|e|$/\AA,
or $-$0.023~Debye/\AA$^2$.  The negative sign means that the dipole points
into Li$_3$PO$_4$ from Au (111).  Using 
\begin{equation}
\Delta {\cal V}_e=4\pi \delta/(|e|) , \label{delta}
\end{equation}
where all quantities are in atomic units, this ``small'' $\delta$ should
yield a large $-$0.86~V shift between bare and phosphate-coated Au(111).
This explains the $-$0.88~V difference deduced from work function
differences discussed above.  DFT calculations give an aggregate $\delta$,
that includes screening, induced-dipoles, and depolarization effects, not
individual contributions.  Hence no effective dielectric constant is present
in Eq.~\ref{delta}.\cite{tung2} 

Under strictly UHV conditions, no net charge can be introduced.  Net bulk or
surface charges lead to infinite repulsive energies due to long-ranged
coulombic repulsion.  Instead, we raise the voltage by removing matched
Li$^+$/$e^-$ pairs, as is normally done in DFT calculations.\cite{zhou,greeley}
The configuration space associated with removing multiple Li is large.  We
assume that Li vacancies are weakly interacting, and their formation energetics
only depend on their positions in the $z$ direction, perpendicular to the
surface.  Vacancies on the same layer of Li$_3$PO$_4$
are placed as far apart from each other as possible.  With $\Delta N$(Li)=$-4$,
two different configurations with all Li-vacancies right at the interface are
indeed found to yield total energies within 30~meV of each other.

\begin{figure}
\centerline{\hbox{ \epsfxsize=1.10in \epsfbox{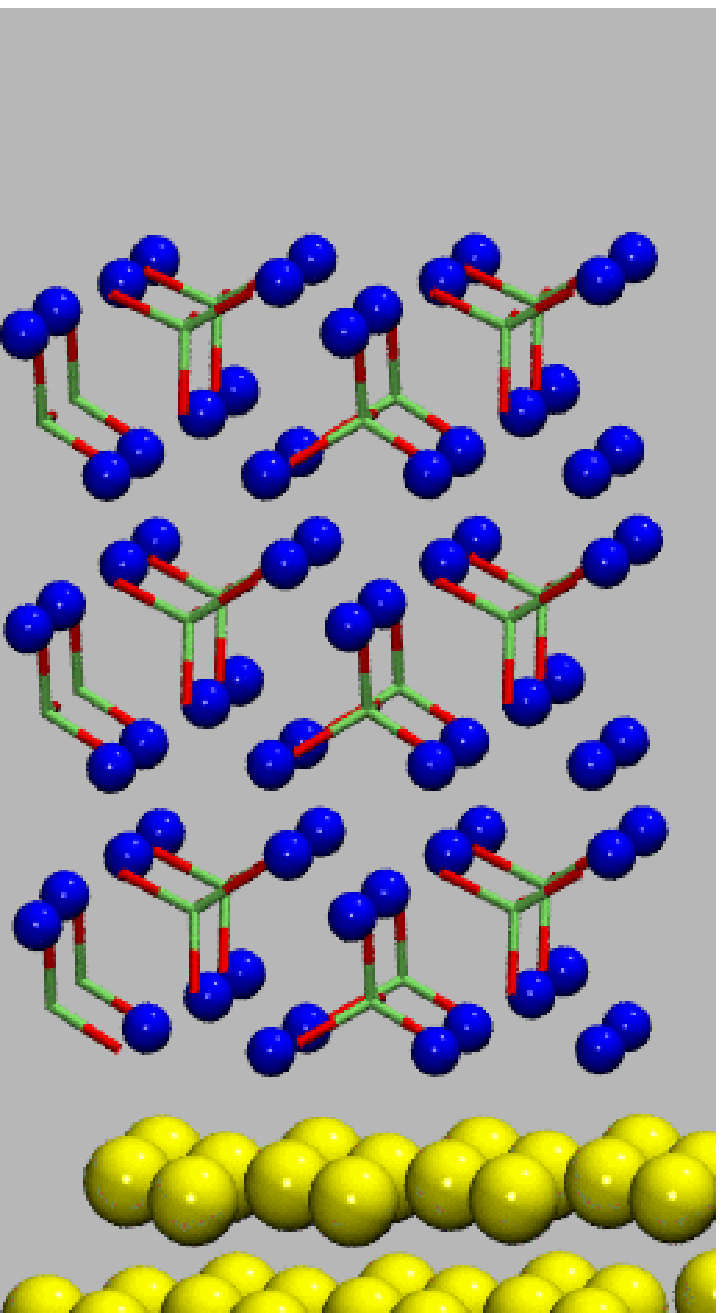} }
            \hbox{ \epsfxsize=1.10in \epsfbox{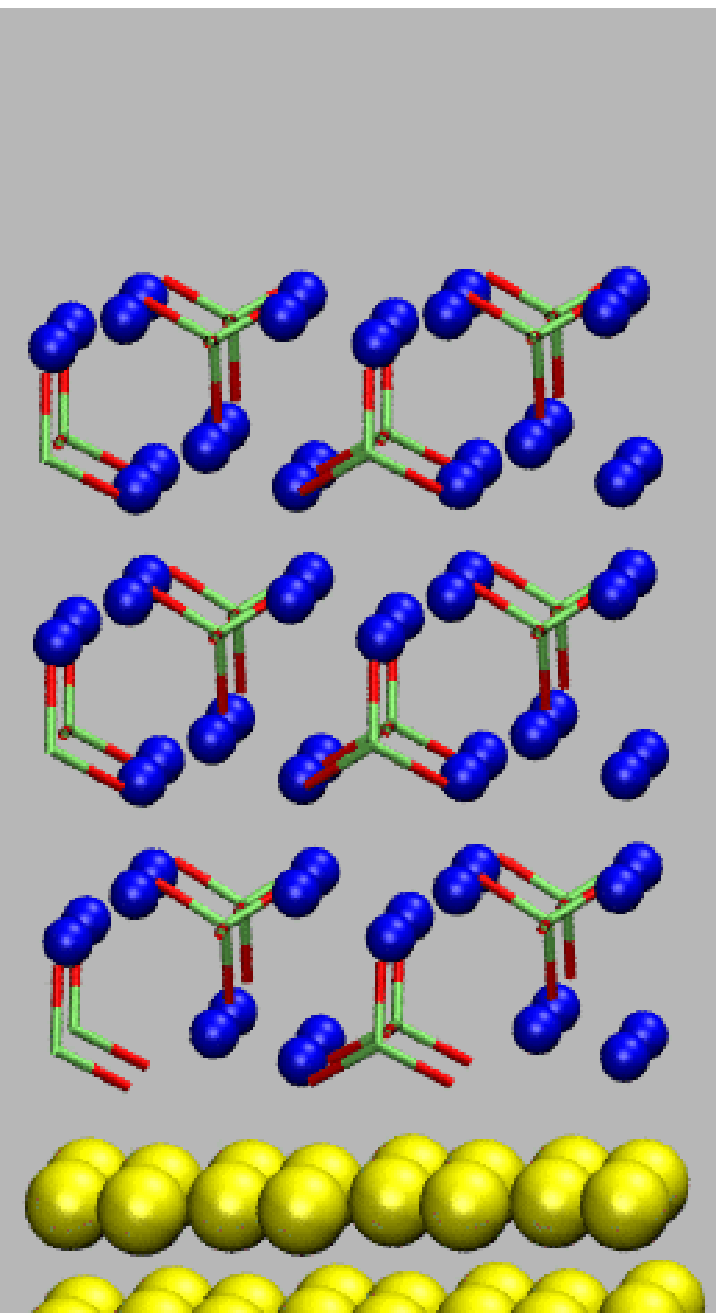} }
            \hbox{ \epsfxsize=1.10in \epsfbox{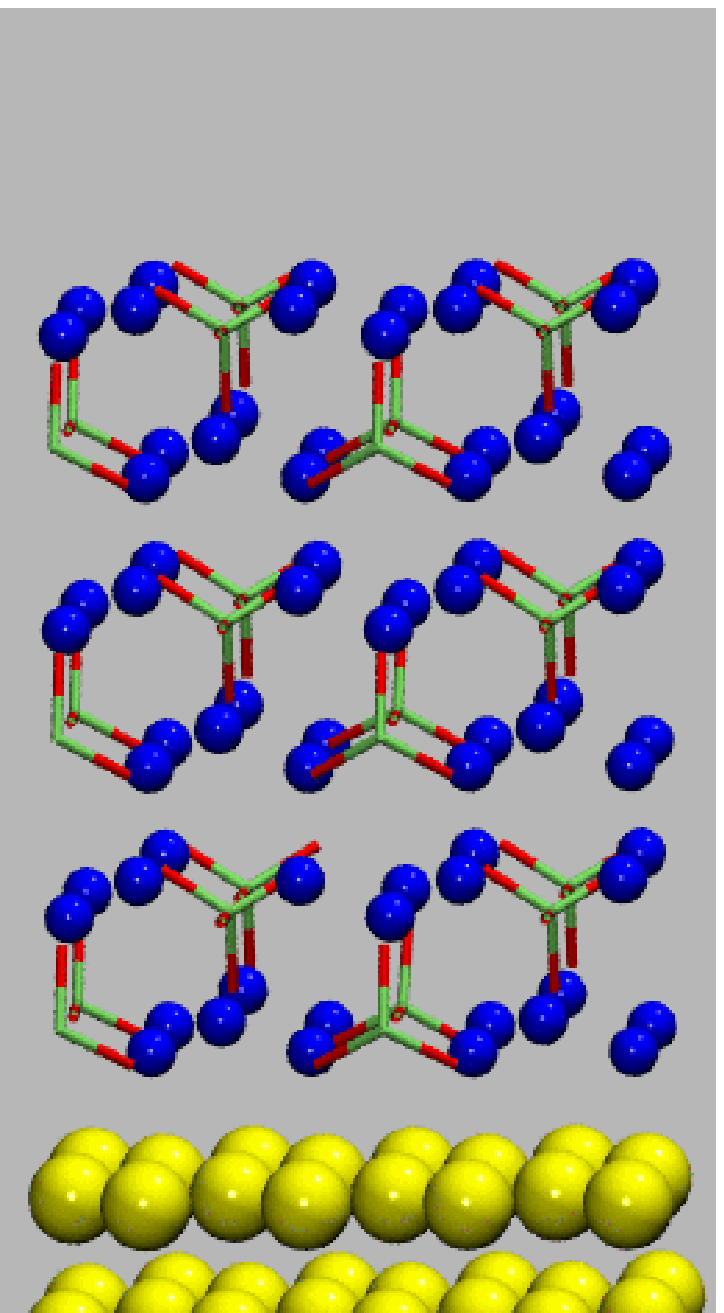} }}
\centerline{\hbox{ (a) \hspace*{0.80in} (b) \hspace*{0.8in} (c)}}
\caption[]
{\label{fig5} \noindent
Li$_{3-x}$PO$_4$/Au: $\Delta N$(Li)= (a): -1; (b): -4; (c): -1$^a$.
Panel (c) (case $a$) is associated with a Li vacancy in the middle
of the phosphate slab.  See Fig.~\ref{fig2} caption for color key.
}
\end{figure}

Removing one Li (i.e., $\Delta N$(Li)=$-1$) from the Li$_3$PO$_4$ layer in
contact with Au is most energetically favorable (Fig.~\ref{fig5}a).  It
also gives the lowest ${\cal V}_e$ at this $n_{\rm Li}$ content.  The DOS there
reveals no charge transfer between Au and Li$_3$PO$_4$ (Fig.~\ref{fig4}b).  
Some spurious Li$_3$PO$_4$ ``orbitals'' in contact with Au(111) are above
$-1.0$~eV, but they likely arise from the arbitary spatial decomposition
scheme; the valence band edge should be taken as $-1$~eV where at least
two layers of Li$_3$PO$_4$ atoms exhibit the same orbital level.\cite{note7}
A net negative charge must remain at the Li-vacancy site,
compensated by a positive charge on Au(111) (Fig.~\ref{fig1}e).  This charge
separation modifies $\delta$ to $-$0.0029~$|e|$/\AA\, which should yield a
voltage shift of $-0.53$~V from the bare Au(111) value.  Using explicit $\Phi$
calculations, ${\cal V}_e$ is found to rise to 3.28~V, shifted from the Au(111)
${\cal V}_e$ by $-$0.52~V (Table~\ref{table1}), which is quantatitively
explained by the change in $\delta$.  The phosphate valence band edge is raised
by about 0.45~eV relative with $E_{\rm F}$ on the Au surface (Fig.~\ref{fig4}b).
Removing more Li at the phosphate-gold interface augments $\delta$ in the
positive direction, increases ${\cal V}_e$ (Table~\ref{table1}), and raises
the valence band edge proportionately (Fig.~\ref{fig4}c).  Finally, removing
a single Li inside the phosphate slab (Fig.~\ref{fig4}d) or at the
phosphate-vacuum interface (Fig.~\ref{fig4}e) yields a larger ${\cal V}_e$
than if the vacancy is right at the Au(111) surface (Table~\ref{table1}).
In the latter case, the phosphate valence band edge now coincides with
$E_{\rm F}$.

So far we have not discussed the ionic voltage (${\cal V}_i$) governing the
Li content.  Table~\ref{table1} shows that ${\cal V}_e$ exhibits a
significantly larger slope as a function of $\Delta N$(Li) than ${\cal V}_i$
as long as Li are removed from the interior interface.  The two voltage
definitions should intersect at only one point.  ${\cal V}_e$ approaches
${\cal V}_i$ between $\Delta N$(Li)=$-1$ to $-2$.  Away from that solid-state
PZC, equilibrium (${\cal V}_e$=${\cal V}_i$) can be achieved with a net surface
charge compensated by liquid electrolytes (Fig.~\ref{fig1}g,
Sec.~\ref{li2co3_ec}), just like on noble metal electrodes.  Models with
non-equilibrium interfacial Li content are ``wrong'' in the sense that they
only have transient existence.  

Removing a Li sitting deeper inside the phosphate layer
(Fig.~\ref{fig1}f,~\ref{fig5}b) shows signs of yielding self-consistency:
${\cal V}_e$=3.78~V, ${\cal V}_i$=3.69~V (Table~\ref{table1}).  However, the
relative free energy ($\Delta G)$ of this alternatve $\Delta N$(Li)=$-1$
configuration, evaluated using
\begin{equation}
\Delta G = G(\Delta N({\rm Li}))-G(0) -\Delta N({\rm Li}) \mu_{\rm Li} \, ,
\label{deltag}
\end{equation}
is less favorable than $\Delta N$(Li)=$-3$ where all Li vacancies are at the
interface (Table~\ref{table1}, fifth column), at a similar Li chemical
potential ($\mu_{\rm Li}$=${\cal V}_i$).  In general, if
${\cal V}_i$ increases upon successive Li removal right at the interface,
Eq.~\ref{deltag} favors removing multiple Li there, rather than removing a
single Li further inside the slab -- even if the latter yields the same
${\cal V}_i$.  Thus Fig.~\ref{fig1}f should not be a candidate for 
equilibrium interfaces,\cite{rossmeisl15} at least under these UHV conditions.
If a liquid electrolyte is present, it cannot be ruled out that Li$^+$ vacancy 
near the outer surface may be stabilized; we speculate that this could even
yield a multitude of equilibrium Li$^+$ configurations in Li-containing
films, and therefore more than one PZC at different voltages.  

\subsection{Electrostatics-based Passivation Strategy} \label{strategy}

These results emphasize that the interfacial dipole density ($\delta$) is one
atomic lengthscale determinant of applied voltages.  From Eq.~\ref{delta},
a thin dipole sheet is expected to give rise to a step-like voltage drop at
the interior interface of Li$_3$PO$_4$ (Fig.~\ref{fig1}e), which is
therefore not an inert, passive coating (Fig.~\ref{fig1}d).  At long enough
times such that Li$^+$ can insert and relax their configurations, this solid
electrolyte should absorb much of the voltage drop/rise at the sharp
metal/electrolyte interface.  If a liquid electrolyte exists in the vacuum
region of Fig.~\ref{fig5}, it should experience a reduction in the extreme
potential exerted by the cathode (Fig.~\ref{fig1}e), in effect widening the
liquid stability window.  Solid films coated on anodes should work analogously;
in that case, the film, or at least its interface with the anode, must be able
to incorporate extra Li.\cite{maier}

The above discussions suggest a new design principle for ``smart'' artificial
SEI passivating lithium ion battery anodes and high-voltage cathodes.
Traditional SEI relies on Li$^+$-conducting but $e^-$ insulating films
to slow down $e^-$ tunneling from electrodes to organic-solvent based
electrolytes.  They are not designed to affect the voltage experienced by
the liquid electrolyte, which thermodynamically speaking should still
decompose at battery operating conditions.  Our proposed scheme focuses on
local thermodynamics, not $e^-$ tunneling kinetics.

A fast Li$^+$ transport rate is crucial for successful implementation of 
passivation based on Li$^+$-motion induced interfacial dipoles.  Another
route, not involving Li motion, is to incorporate permanent dipoles that can
re-orient, such as ferroelectric films (Fig.~\ref{fig1}b).  Ferroelectric
surfaces have already been used to manipulate photochemical
reactivities.\cite{ferro1} Most ferrorelectrics are not Li$^+$ conductors.
However, electrostatic effects are long-ranged, and it is likely that
non-continuous ferrorelectric films, lined with gaps that Li$^+$ can pass
through, can serve to reduce extreme voltages experienced by liquid
electrolytes.

\subsection{Au/Li$_2$CO$_3$/vacuum (Fig.~\ref{fig3}b) for Summary}\label{li2co3}

\begin{table}\centering
\begin{tabular}{ |c|r|r|r|r|r|r| } \hline
$\Delta N$(Li)  & 0 & -1 & -2 & -3 & -4 & -6 \\ \hline
${\cal V}_e$ & 3.56 & 3.87 & 4.03 & 4.03 & 4.05 & 4.41 \\
${\cal V}_i$ & 3.53 & 3.58 & 3.26 & 3.55 & NA & NA \\ \hline
\end{tabular}
\caption[]
{\label{table2} \noindent
Electronic (${\cal V}_e$) and ionic (${\cal V}_i$) voltages for a Li$_2$CO$_3$
basal plae/Au(111) interface as the number of Li vacancies at the inner
surface of the Li$_2$CO$_3$ slab varies.  No adsorbed molecule exists.
}
\end{table}

The band alignment between Au(111) and the stoichiometric Li$_2$CO$_3$ basal
plane (Fig.~\ref{fig2}b) is similar to that predicted for Au/Li$_3$PO$_4$.
In the S.I., the Au(111)/Li$_2$CO$_3$(001) interface is also shown to be
qualitatively similar.  The broad correspondence suggests that the predicted
behavior is universal for non-redox-active films.

\begin{figure}
\centerline{\hbox{ \epsfxsize=3.50in \epsfbox{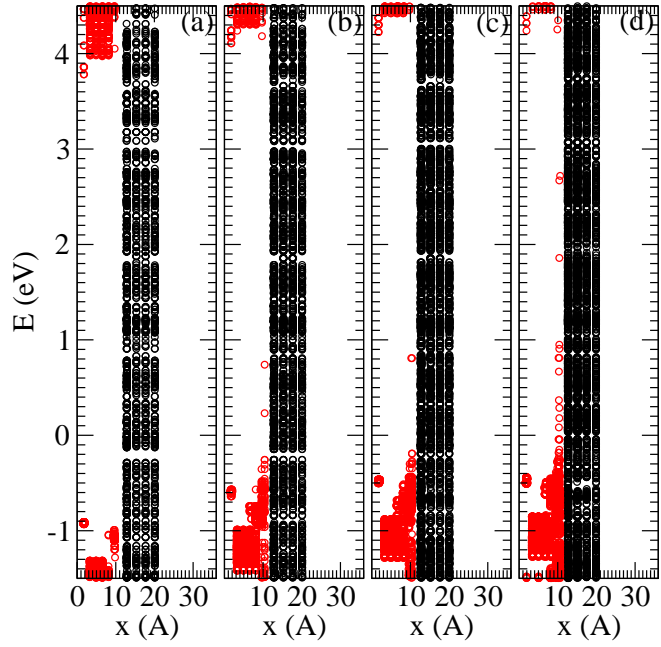} }}
\caption[]
{\label{fig6} \noindent
Kohn-Sham orbitals of Li$_{2-x}$CO$_3$/Au decomposed on to atoms at their
coordinates perpendicular to the interface.  (a): stoichiometric; (b):
2~Li$^+$ vacancies inside; (c): 3~Li$^+$ vacancies inside; (d): 4~Li$^+$
vacanies inside.  $E_{\rm F}$=0.0~eV.  The vacuum levels are at 4.93, 5.40,
5.40, and 5.40~eV in the four panels.  Au and thin-film-based orbitals are
depicted in black and red, respectively.  $\rho_c$=0.005.\cite{note7}
}
\end{figure}

The valance and conduction band edges of the $\Delta N$(Li)=0 basal Li$_2$CO$_3$
slab bracket the Au Fermi level (Fig.~\ref{fig6}a), indicating no $e^-$
transfer.  ${\cal V}_i$ associated with this Li$_2$CO$_3$ slab is predicted to
be 3.53~V, very close to ${\cal V}_e$=3.58~V there.   Thus the PZC somewhat
fortuitously occurs at stoichiometry.  Next, we raise the voltage and
try to oxidize Li$_2$CO$_3$.\cite{saito2011,liair1} by removing successive
Li$^+$/$e^-$ pairs.  The most favorable Li~vacancies are at the
Au/carbonate interface (Fig.~\ref{fig7}a-b).  As Li are further removed
(Fig.~\ref{fig7}c, Fig.~\ref{fig6}b-c), ${\cal V}_e$ increases
(Table~\ref{table2}).  The valence band also rises until the occupied states
at the Au/carbonate interface reaches $E_{\rm F}$ (Fig.~\ref{fig6}d).  At this
point, $e^-$ transfers from Li$_2$CO$_3$ to Au(111).  Oxidation under UHV
conditions thus initiate from the Au/Li$_2$CO$_3$ interface.  The redox
reaction pins ${\cal V}_e$, which ceases to increase with another Li removal
(Table~\ref{table2}).  This is reminiscent of ``Fermi level pinning'' at
metal-semiconductor interfaces by surface/defect states,\cite{tung1} except
that in LIB the ``defects'' are mobile Li$^+$ vacancies, not static impurities,
and their density can be much higher.  When 6 Li vacancies are created at the
interface, ${\cal V}_e$ increases again.  This may due to the complete removal
of $e^-$ from some surface states.  Unlike Li$_3$PO$_4$, the interfacial layer
of Li$_2$CO$_3$ starts to lose structural integrity with removal of multiple
Li$^+$ (Fig.~\ref{fig7}c).

Removing one Li from the middle of the Li$_2$CO$_3$ slab (the hole remains on
the Au surface) is less favorable by 0.25~eV compared to removing Li$^+$ at
the Au/carbonate interface.  Removing a Li$^+$ on the vacuum surface
(Fig.~\ref{fig7}d) is less favorable by another 0.22~eV, and
${\cal V}_e$=4.50~V, much higher than the most energetically favorable Li$^+$
vacancy at the interior Au(111) interface.  These configurations reveal an
inhomogeneity in the energy landscape perpendicular to the interface that may
strongly affect Li$^+$ transport\cite{santosh} -- especially if Li$^+$ has to
escape from the interior interface to the outer carbonate surface to the liquid
electrolyte.

\begin{figure}
\caption[]
{\label{fig7} \noindent
Li$_{2-x}$CO$_3$/Au: (a): 1~Li$^+$ vacancy inside; (b): 2~Li$^+$ vacancies
inside; (c): 4~Li$^+$ vacancies inside; (d) 1~Li$^+$ vacancy outside.  (e) 6~EC
on stoichiometric Li$_2$CO$_3$/Au; (f) 5~EC, 1~PF$_6^-$ on Li$_2$CO$_3$/Au.
F is in pink; see Fig.~\ref{fig2} caption for color key.  (The figures
are removed due to file size limitations -- they can be form in the
JPCC paper or are available upon requres.)
}
\end{figure}

\subsection{Au/Li$_2$CO$_3$/EC/vacuum (Fig.~\ref{fig3}c)for summary}
\label{li2co3_ec}

Next we use this Li$_2$CO$_3$ example to illustrate the impact of including
explicit molecules in the theoretical models.  A monolayer of adsorbed ethylene
carbonate (EC) molecules is shown to exhibit profound effect on predicted
voltages.   Such molecular adsorption might be realized under UHV settings.
We stress that frozen EC layers at $T=0$~K cannot be considered adequate
approximations of liquids at finite temperature.\cite{interface}

\begin{figure}
\centerline{\hbox{ \epsfxsize=3.00in \epsfbox{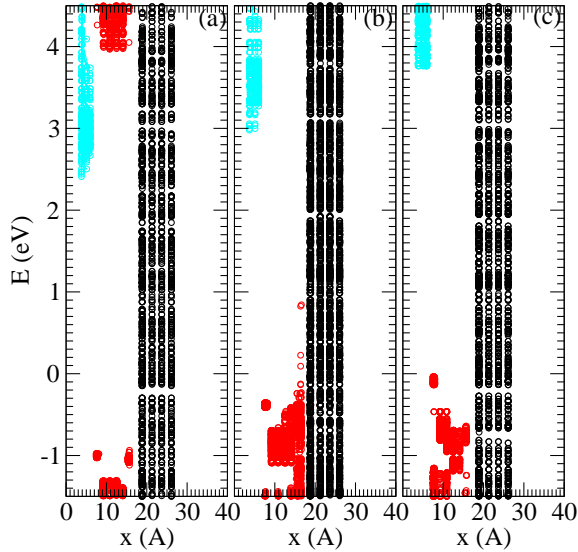} }}
\caption[]
{\label{fig8} \noindent
Kohn-Sham orbitals of EC/Li$_{2-x}$CO$_3$/Au decomposed on to atoms at
their coordinates perpendicular to the interface.  (a): stoichiometric;
(b): 2~Li$^+$ vacancies inside; (c): a PF$_6^-$ replacing one EC in panel
(a).  The vacuum levels are at 3.12, 3.68, and 4.91~eV in the three panels.
Au, Li$_2$CO$_3$, and EC-/PF$_6^-$-based orbitals are depicted
in black, red, and blue, respectively.  $\rho_c$=0.005.\cite{note7}
}
\end{figure}

Fig.~\ref{fig7}e corresponds to Fig.~\ref{fig7}a except for
the adsorbed EC monolayer.  The solid film atomic structure,
energetics of successive Li removal (${\cal V}_i$, Tables~\ref{table2}
and~\ref{table3}), and DOS (Fig.~\ref{fig6}a and Fig.~\ref{fig8}a, after
shifting $E_{\rm F}$ to zero), are almost identical with or without EC.
This is unsurprising because the EC layer neither transfers $e^-$ to nor
form a chemical bond with Li$_2$CO$_3$.  However, this ``spectator'' EC layer
leads to a large ${\cal V}_e$ decrease.  They carry sufficiently large
dipole moments, even when tilted almost flat in the adsorption geometry,
to reduce ${\cal V}_e$ by $>$1~V compared to vacuum
interfaces.\cite{delta_notes}  At liquid-solid interfaces, thermal fluctuations
should modify the $\delta$ value predicted for the monolayer EC in vacuum at
T=0~K.  Nevertheless, liquid EC at finite temperature has also been shown to
change the PZC of the LiC$_6$ edge plane by more than 1~V.\cite{vedge}
EC molecules are highly asymmetric; they bind strongly to cations, including
Li$^+$ ions at material surfaces,\cite{borodin09,voth} but weakly to anionic
species. The resulting dipole layer decreases the vacuum ${\cal V}_e$ on
both anodes and cathodes unless the surfaces are sufficiently highly charged
to force reorientation of the molecules.

The DOS of Fig.~\ref{fig8}b may superficially seem to suggest Li$_2$CO$_3$
oxidation can occur at a very low ${\cal V}_e$=2.31~V.  However, ${\cal V}_e$
predicted with no Li vacancy is already lower than ${\cal V}_i$ computed for
removing one Li.  The correct interpretation of the first 3~columns of
Table~\ref{table3} is that the Li content needs to increase in the direction
of ${\cal V}_e$ until all Li-sites are occupied (Sec.~\ref{voltage}).  It is
inconsistent for (Li$^+$,$e^-$) pairs to be removed from the stoichiometric
Li$_2$CO$_3$ slab when ${\cal V}_e$=2.31~V.  Therefore no oxidation occurs.

\begin{table}\centering
\begin{tabular}{ |c|r|r|r|r|r|r|} \hline
$\Delta N$(Li),$N$(PF$_6^-$)  
	& (0,0) & (-1,0) & (-2,0) & (0,1) & (-1,1)  \\ \hline
${\cal V}_e$ & 1.73 & 2.07 & 2.31 & 3.53 & 3.45 \\
${\cal V}_i$ & 3.47 & 3.61 & NA   & 3.57 & NA   \\ \hline
\end{tabular}
\caption[]
{\label{table3} \noindent
Electronic (${\cal V}_e$) and ionic (${\cal V}_i$) voltages for
the Li$_2$CO$_3$ basal plane/Au(111) interface as the number of Li vacancies
at the inner surface of the Li$_2$CO$_3$ slab varies.  The outer surface
is coated with 6 EC and/or 5 EC and a PF$_6^-$.
}
\end{table}

So far we have considered uncharged electrodes.  If a liquid electrolyte is
present, net surface charges can be compensated by mobile ions in the liquid
electric double layer (Fig.~\ref{fig1}g).  Fig.~\ref{fig7}f and~\ref{fig8}c
represent a charged interface after replacing an EC molecule with a
PF$_6^-$.\cite{pf6_notes} The simulation cell remains charge neutral, and the
$-|e|$ charge must be largely compensated by a net positive charge on the
Au(111) surface.  This sets up an additional dipole moment, which increases
$\Phi$ (Eq.~\ref{delta}).  As a result, ${\cal V}_e$ rises to 3.53~V, very
close to ${\cal V}_i$ under these conditions (Table~\ref{table3}).  The DOS
(Fig.~\ref{fig8}c) shows that, while the PF$_6^-$ valence band remains well
below $E_{\rm F}$, some Li$_2$CO$_3$ orbitals on the outer surface has been
raised almost to the Fermi level due to the close proximity of the anion.
Although the extent of the surface charge depletion is difficult to quantify,
Fig.~\ref{fig8}c suggests that oxidation may initiate at the outer surface of
Li$_2$CO$_3$ in liquid electrolyte if PF$_6^-$ coordinates to the carbonate
surface in an inner-shell configuration (i.e., they are in physical contact),
and if the carbonate film is thin enough to allow electron transport.

A thicker carbonate film will increase $\delta$ at the same PF$_6^-$ surface
density because of the larger charge separation.  This means that fewer
PF$_6^-$ ion near the surface would be needed to generate the same voltage
increase (Eq.~\ref{delta}).  We speculate that net charges on thick film
surfaces are sparse and can be treated as isolated defects.  The anions may
also be farther away inside the liquid region and be screened by
solvent molecules.  As a result, the center of negative charge may or may not
become more distant from the surface, and a quantative estimate of the effect
of anion-surface separation on the voltage needs to be explicitly computed.
Whether thin film-coated electrodes exhibit net charges compensated by
counterions in the liquid (Fig.~\ref{fig1}g), or whether the voltage increase
can be accommodated entirely by changes in Li$^+$ configurations in the solid
region (Fig.~\ref{fig1}f), depends on which option minimizes the Gibbs free
energy.

\subsection{Au/LiMn$_2$O$_4$/vacuum and Au/LiMn$_2$O$_4$/EC/vacuum
	(Fig.~\ref{fig3}d for Summary)} \label{mno}

In previous work, we have considered EC oxidative decomposition
on Li$_x$Mn$_2$O$_4$ (100)\cite{lmo1} and (111)\cite{lmo2} surfaces, in both
UHV conditions\cite{lmo1,lmo2} and in explicit liquid electrolyte.\cite{lmo1}
The liquid is found not to strongly affect the rate-determining
reaction barrier on the (100) surface.  The voltage dependence of EC
decomposition was not considered there because Li$_x$Mn$_2$O$_4$, like most
transition metal oxides, does not have a readily-defined $E_{\rm F}$.

In this work, $E_{\rm F}$ is made unambiguous by putting a thin,
reconstructed\cite{persson} LMO (111) slab on a Au (111) ``current collector''
(Fig.~\ref{fig2}c).  Removing an $e^-$ from the system, without removing
an accompanying Li$^+$, does not immediately lead to a change in any Mn
ion charge state, strongly suggesting that the $e^-$ is removed from the
Fermi level straddling Au orbitals (not shown).  This confirms that $E_{\rm F}$
is relevant for voltage estimate via Eq.~\ref{instant}.  Allowing geometry
optimization after $e^-$ removal yields a localized hole polaron which is
now in equilibrium with $E_{\rm F}$.

\begin{figure}
\centerline{\hbox{ \epsfxsize=1.00in \epsfbox{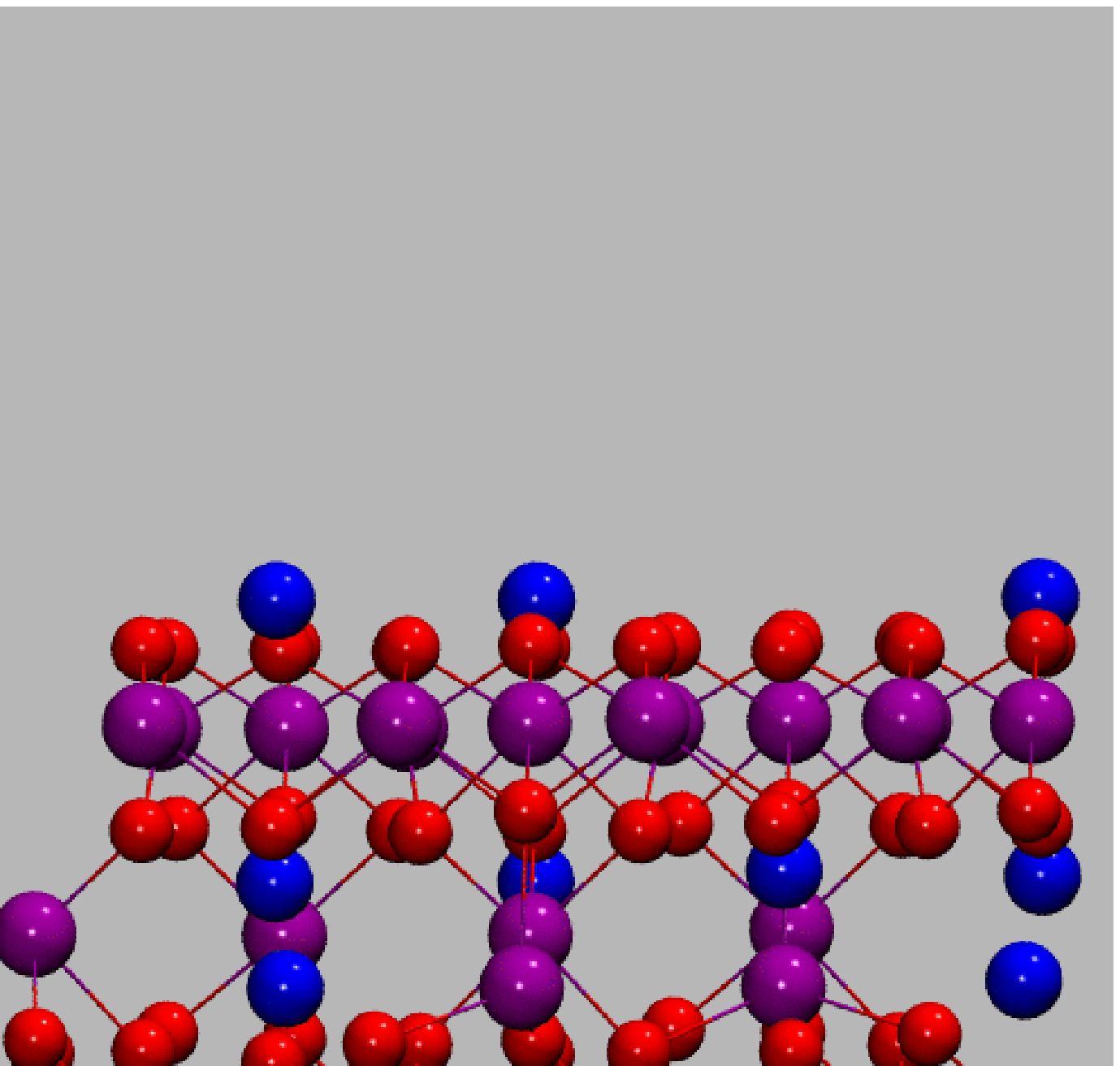} }
            \hbox{ \epsfxsize=1.00in \epsfbox{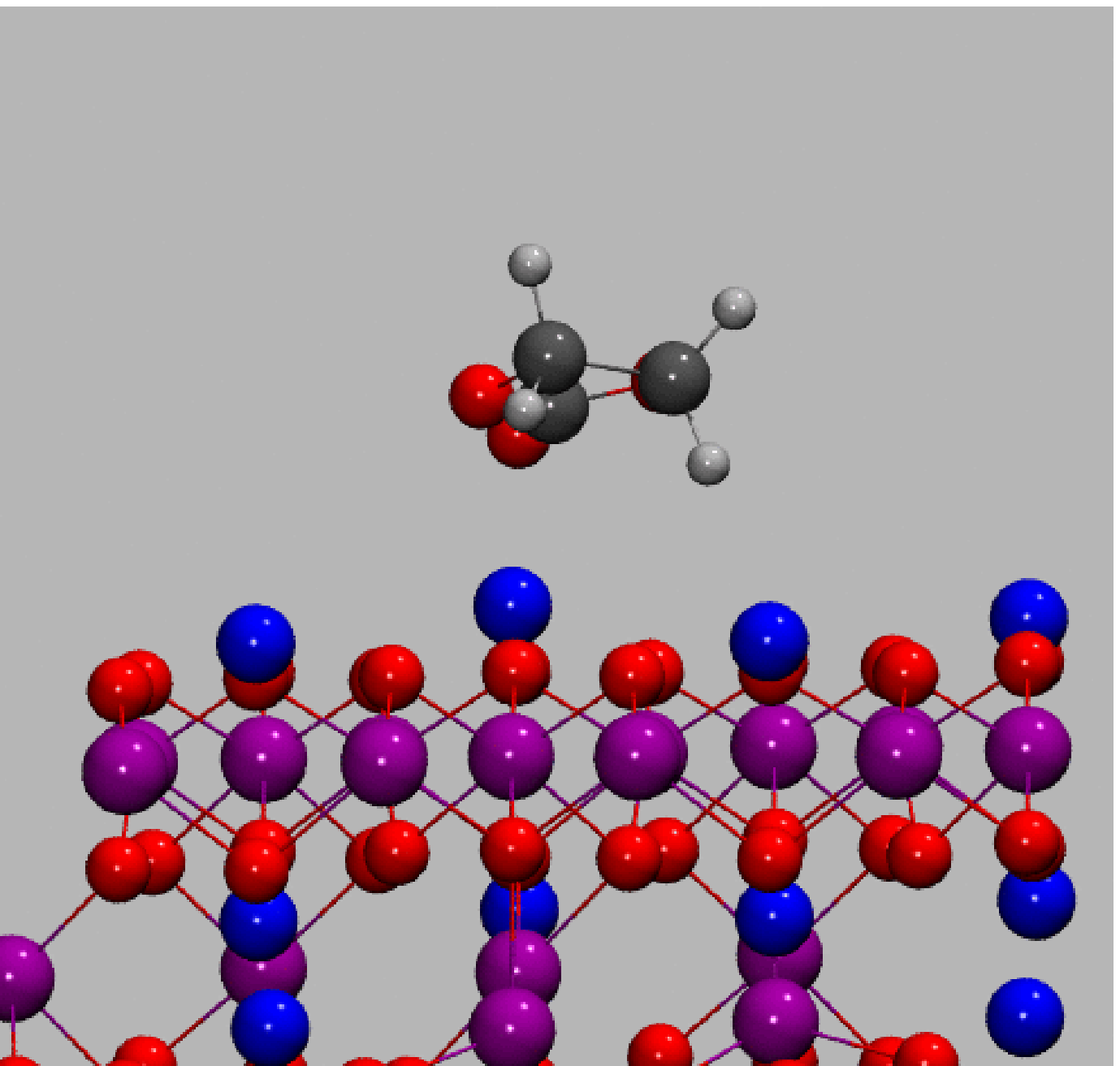} }
	    \hbox{ \epsfxsize=1.00in \epsfbox{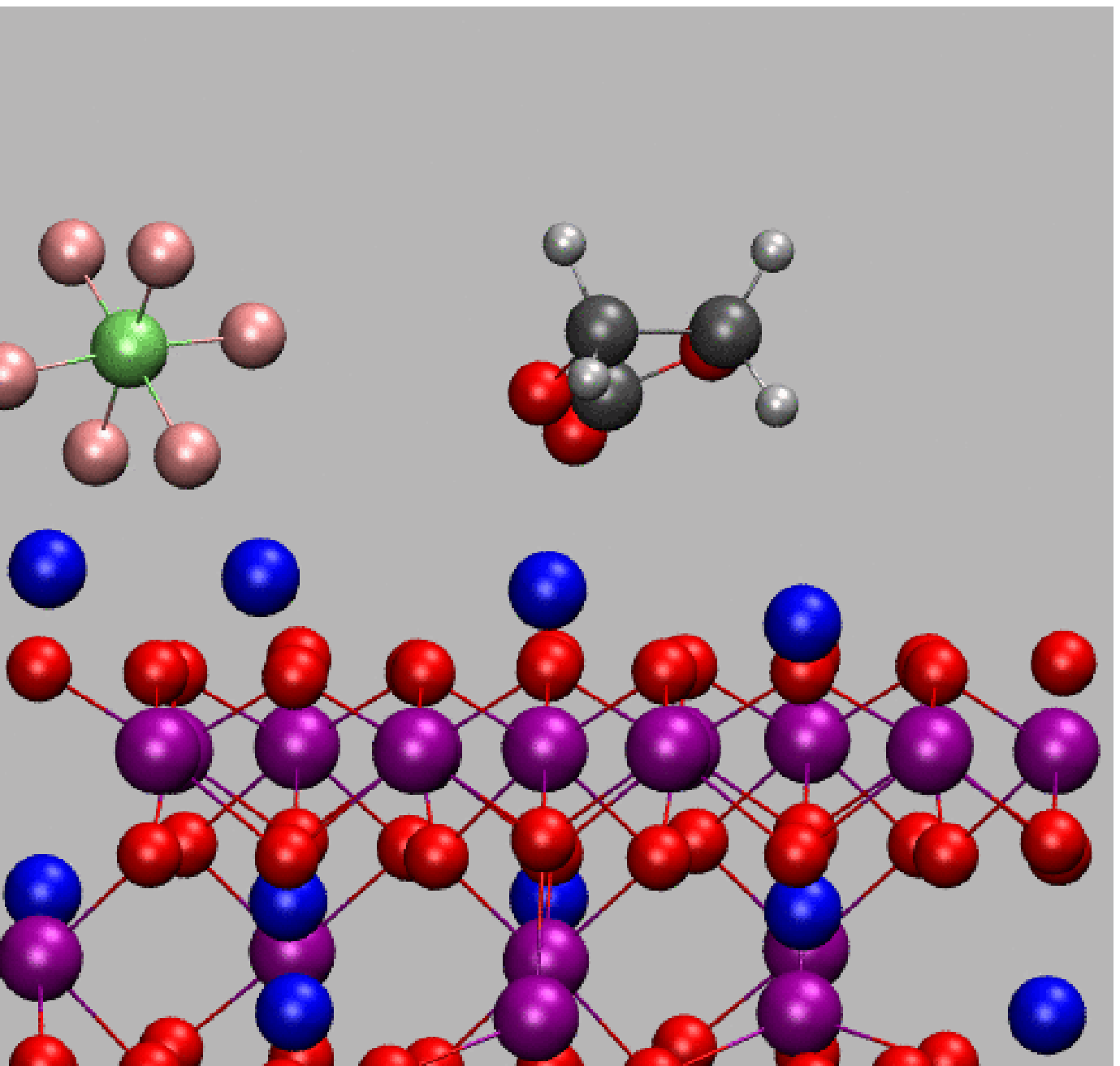} }}
\centerline{\hbox{ (a) \hspace*{0.9in} (b) \hspace*{0.9in} (c)}}
\centerline{\hbox{ \epsfxsize=1.00in \epsfbox{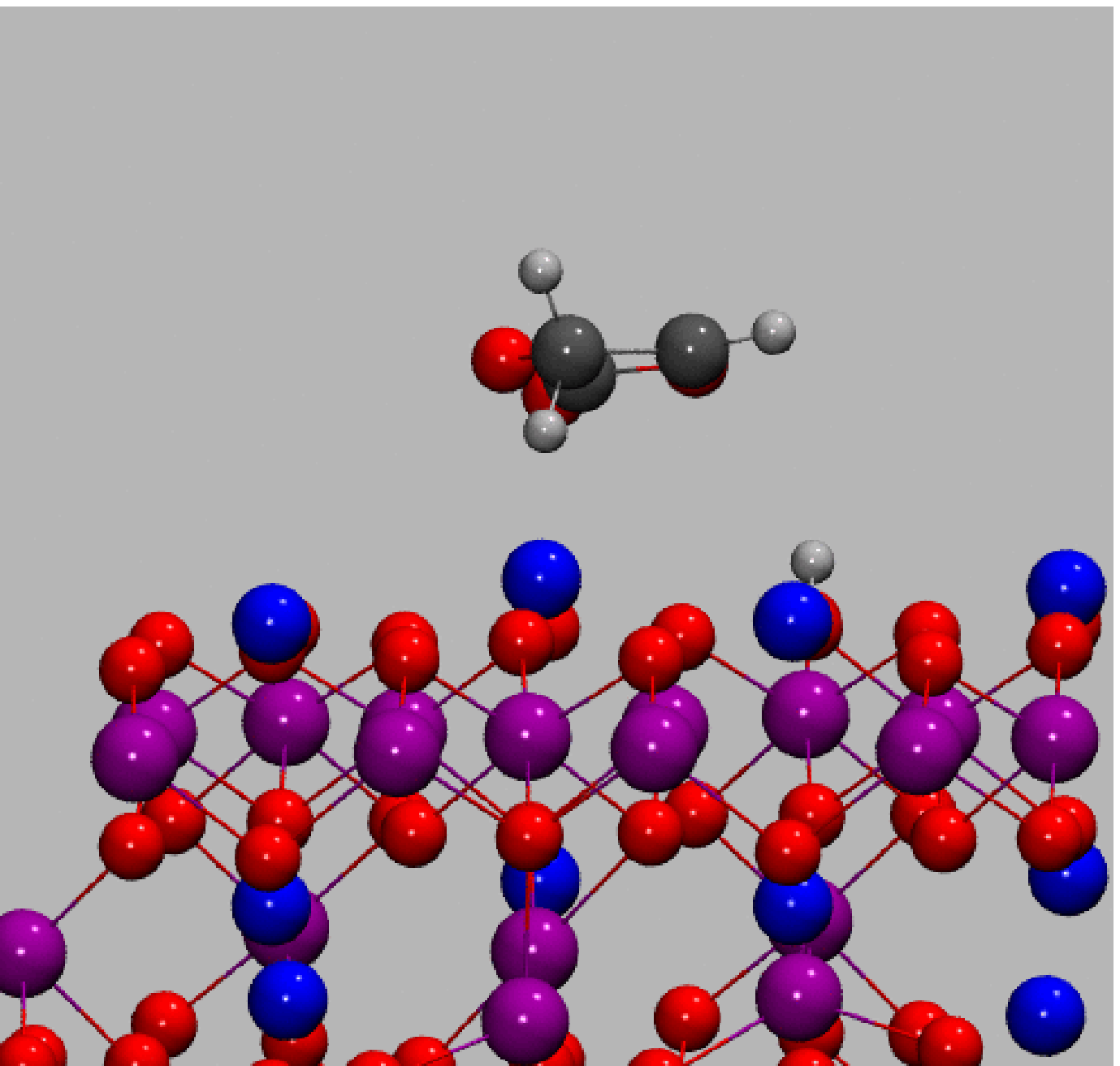} }
            \hbox{ \epsfxsize=1.00in \epsfbox{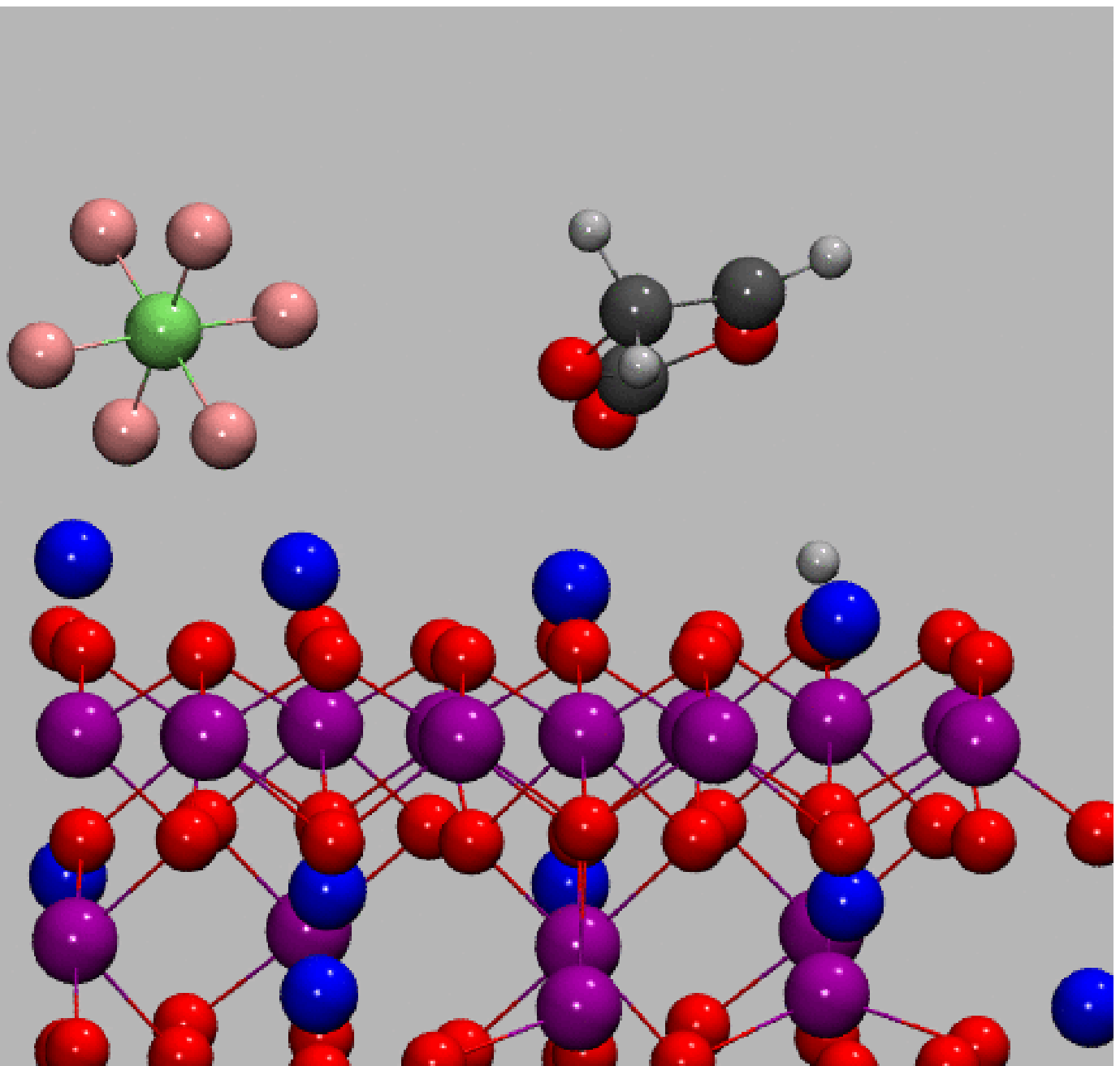} }}
\centerline{\hbox{ (d) \hspace*{0.7in} (e)}}
\caption[]
{\label{fig9} \noindent
EC and PF$_6^-$/Li$_{1-x}$Mn$_2$O$_2$/Au.  Au atoms are not shown.
(a) $\Delta N$(Li)=-5; 
(b) $\Delta N$(Li)=-4; with 1 intact EC;
(c) $\Delta N$(Li)=-4; with 1 intact EC and 1~PF$_6^-$;
(d) $\Delta N$(Li)=-4; with 1 broken EC;
(e) $\Delta N$(Li)=-4; with 1 broken EC and 1~PF$_6^-$.
$\Delta N$(Li) is the number of surface Li vacancies compared to the
stoichiometric slab.  See Fig.~\ref{fig2} caption for color key.
}
\end{figure}

In contrast to Li$_2$CO$_3$ and Li$_3$PO$_4$, successive removal of
Li$^+$/$e^-$ pairs from the stoichiometric LMO slab occur most favorably from
the outer surface (Fig.~\ref{fig9}a, see also the S.I.), not the LMO/Au
interface.  A Mn(III) turns into a Mn(IV) to accompany each $e^-$ removal.
Table~\ref{table4} indicates that the voltage increases with decreasing
surface Li content, and ${\cal V}_e$=${\cal V}_i$ is achieved
at $\Delta N$(Li)=$-$2 under UHV conditions.

\begin{table}\centering
\begin{tabular}{ |c|r|r|r|r|} \hline
$\Delta N$(Li),$N$(EC),$N$(PF$_6^-$)
 & (-2,0,0) & (-3,0,0) & (-4,0,0) & (-5,0,0) \\ \hline
${\cal V}_e$ & 3.28 & 4.33 & 4.71 & 5.24 \\
${\cal V}_i$ & 3.27 & 3.58 & 4.40 & NA   \\ \hline
$\Delta N$(Li),$N$(EC),$N$(PF$_6^-$)
 & (-3,1,0) & (-4,1,0) & (-5,1,0) & (-4,1,1)\\ \hline
${\cal V}_e$ & 4.02 & 4.23 & 4.78 & 6.39 \\
${\cal V}_i$ & 3.86 & 4.19 & NA & NA \\ \hline
\end{tabular}
\caption[]
{\label{table4} \noindent
Electronic (${\cal V}_e$) and ionic (${\cal V}_i$) voltages for the
Li$_x$Mn$_2$O$_4$(0001)/Au(111) interface as the number of Li vacancies at
the outer surface varies.
}
\end{table}

Next we add EC to the LMO surface. If the system were a liquid-solid interface,
the surface Li content, net surface charge, and surface density of EC molecules
which directly coordinate to available Li ions should be self-consistently 
deduced using AIMD simulations at finite temperature.\cite{interface}  Here we
confine ourselves to $T=0$~K and to adding one EC, yielding the surface used
in Ref.~\onlinecite{lmo2}.  $\Delta N$(Li)=-4~in this model (Fig.~\ref{fig9}b),
and ${\cal V}_e$=4.23~V, in reasonably agreement with ${\cal V}_i$=4.19~V for
this configuration (Table~\ref{table4}).  The voltage is within the
experimental LMO operating range.  Except for the shift in ${\cal V}_e$, and
hence in the vacuum level, the DOS is almost unchanged from the case where no
EC is present (Fig.~\ref{fig10}a-b).  With no anion on the surface, the EDL
structure should resemble Fig.~\ref{fig1}a.

It is also important to elucidate the EDL structure of a charged,
polaron-conducting LMO surface, not considered previously.\cite{lmo2}  As
in the Li$_2$CO$_3$ example, we impose a positive charge by optimizing the
geometry of a PF$_6^-$ anion on the LMO surface in the overal charge-neutral
simulation cell (Fig.~\ref{fig9}c).  The DOS (Fig.~\ref{fig10}c) reveals that
PF$_6^-$ retains its excess electron and is not oxidized.\cite{pf6_notes} A
net $+|e|$ compensating charge is found in the LMO slab, on a second
layer Mn ion (``Mn 14'') below the LMO/vacuum interface.  This layer contains
Mn(II) ions, exchanged with Li$^+$ in the Persson
group's reconstruction,\cite{persson} which may be the reason a Mn(III)
ion there can lose an $e^-$ most readily.  The dipole moment created by the
Mn(IV)-PF$_6^-$ charge separation raises ${\cal V}_e$ dramatically, from
4.25~V to 6.41~V.  Comparing the DOS with (Fig.~\ref{fig10}b) and without
(Fig.~\ref{fig10}c) the PF$_6^-$, an occupied $d$-orbital on ``Mn 14'' is
indeed seen to be depopulated; its orbital energy level now moves above
$E_{\rm F}$.  The EDL in this redox-active material now resembles
Fig.~\ref{fig1}h.  A lower surface anion density will yield smaller shifts
${\cal V}_e$ but will require a simulation cell with a much larger lateral
surface area.  

\subsection{EC decomposition on LiMn$_2$O$_4$/Au surface}
\label{ec_decomp}

Finally, we re-examine EC decomposition on the $\Delta N$(Li)=$-$4 LMO slab,
now with Au(111) present.  Fig.~\ref{fig9}d depicts a partially decomposed
EC molecule on this LMO surface.  A H$^+$ is transferred to the LMO slab
without breaking any C-O bond in the EC.  Its energy is +0.35~eV above the
intact EC configuration (Fig.~\ref{fig9}b), in good agreement with that of
the reaction intermediate depicted in Fig.~5c of Ref.~\onlinecite{lmo2} where
the Au slab was absent.  A subsequent, C-O bond-breaking step is needed to
render EC oxidation favorable.\cite{lmo2}

\begin{figure}
\centerline{\hbox{ \epsfxsize=4.50in \epsfbox{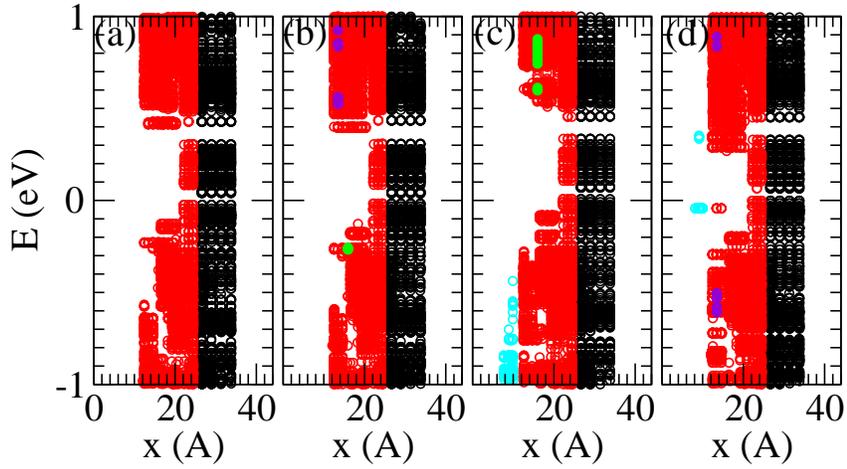} }}
\caption[]
{\label{fig10} \noindent
Kohn-Sham orbitals of EC and PF$_6^-$/Li$_{1-x}$Mn$_2$O$_4$/Au,
decomposed on to atoms at their coordinates perpendicular to the interface.
(a) No adsorbed species.
(b) 1~intact EC;
(c) 1~intact EC and 1~PF$_6^-$.  
(d) 1~broken EC.
$\Delta N$(Li)=-4 in all cases.  
The vacuum levels are at 6.08, 5.60, 7.75, and 4.86~eV in the four panels.
Au, LMO, and EC-/PF$_6^-$-based orbitals are depicted in black,
red, and cyan, respectively.  The filled green and violet circles refer
to $d$-orbitals localized on Mn~14 and Mn~16 (see text).
$\rho_c$=0.005.\cite{note7}
}
\end{figure}

Examining the spin states reveals that a Mn(IV) ion (``Mn 16'') immediately
below the oxide surface gains an $e^-$ to become a Mn(III).  Consistent with
this analysis, the DOS shows an unoccupied Mn $d$-orbital (Fig.~\ref{fig10}b
for intact EC) becoming occupied (Fig.~\ref{fig10}d for broken EC), moving from
above $E_{\rm F}$ to below it due to polaronic relaxation.\cite{mno_notes}
Some orbitals on the decomposed EC fragment reside near $E_{\rm F}$, suggesting
further oxidation can readily occur.

Regardless of whether an $e^-$ is added to LMO from an oxidized EC, or is
removed by imposing a $+|e|$ charge due to a counter-anion nearby, only Mn
redox states change.  Charge localization on Mn ions, accompanied by polaronic
relaxation and a compensating PF$_6^-$ or EC$^+$ fragment nearby, ensures that
$e^-$ and hole never access the Fermi level which lies on Au orbitals
(Fig.~\ref{fig10}).  The Au current collector only serves to establish
$E_{\rm F}$, with which polaron formation in LMO must remain in equilibrium.
Therefore the effect of the applied voltage is indirect.  Increasing
${\cal V}_i$ and ${\cal V}_e$ consistently requires the loss of surface
Li$^+$ (Table~\ref{table4}), which leaves the surface with more oxygen ions
that are under-coordinated, and increases its reactivity towards EC molecules.  
Adsorbed PF$_6^-$ ions also increases the voltage.

Since ${\cal V}_i$ controls the Li content at equilibrium in this voltage range
(Li-sites are not fully depopulated yet), one can arguably ignore both the Au
slab and ${\cal V}_e$ while performing simulations of parasitic reaction on such
redox-active surfaces.\cite{lmo1,lmo2}  As discussed above, the same reaction
intermediate energetics is predicted with or without the Au slab.  However,
${\cal V}_e$ also controls the PF$_6^-$ surface density.  Anions in proximity
of the electrode surface constitute spatial inhomogeneities or ``hot spots''
where parasitic or Li$^+$ insertion reactions may preferentially occur.
Indeed, when a PF$_6^-$ is adsorbed near the EC, the deprotonated EC
intermediate (Fig.~\ref{fig9}e) is found to be 0.1~eV more favorable than the
intact EC, instead of 0.35~eV less favorable when the PF$_6^-$ is absent.
Thus voltage-dependent anion adsorption can affect interfacial processes.
If the cathode material is metallic rather than polaronic, the surface charge
is more uniformly distributed,\cite{voltage,vedge} ${\cal V}_e$ is more
relevant, and qualitatively different behavior may be observed.

\section{Conclusions}

Both electrons and Li$^+$ can move in the surface films covering the electrodes
in lithium ion battery (LIB), and in the electrodes themselves, in response
to voltage variations.  This makes LIB rather unique among electrochemical
devices.  The structures of electric double layers (EDL), and the atomic
lengthscale manifestations of voltages, become much more complex than on
pristine noble metal electrodes.

This work considers Au(111) slabs coated with Li-conducting solid electrolytes
or cathode oxide materials under ultra high vacuum (UHV) conditions.  The
applied voltage is shown to be strongly correlated with the surface dipole
density at the interface, reminiscent of photovoltaics studies.  In relatively
electrochemically inert thin films like Li$_3$PO$_4$ and Li$_2$CO$_3$ coated
on Au(111), dipole moments responsible for voltage increase are established
by creating negatively charged Li$^+$ vacancies at the interface between
cathode surfaces and the thin solid films, compensated with positive charges
on the cathode.  The phosphate example highlights the crucial role of dipole
density.  The carbonate case emphasizes that organic solvent molecules like
ethylene carbonate (EC) can dramatically modify the predicted voltage even
in the absence of charged species (salt).  A corollary is that the
Poisson-Boltzmann theory for liquid electrolytes, which ignores
solvent dipole moments, may not quantitatively describe battery interfaces.

With Li$_x$Mn$_2$O$_4$ (LMO) cathode oxide thin films coated on Au current
collectors, charge transfer occurs at the outer LMO surface and involves
changes in Mn redox states.  The Au slab only indirectly determines the
surface Li content and the anion surface density.  Modeling of electrolyte
decomposition on such electrode surfaces arguably does not require explicitly
including the current collector.  However, the voltage-dependent surface
density of PF$_6^-$ creates inhomogeneities and hot spots where fast
electrolyte decomposition can occur.

We also critically examine the definitions of ``applied voltage'' in DFT
calculations at interfaces.  We distinguish two voltages: electronic
(${\cal V}_e$), due to electronic motion, and ionic (${\cal V}_i$), due to
Li$^+$ redistribution.  In any atomic configuration, ${\cal V}_e$ is well
defined in a metallic electrode and should coincide with the experimental
voltage imposed via a potentiostat.  ${\cal V}_i$ is a self-consistency
critierion.  At equilibrium, the system is pinned by Li$^+$ insertion redox
reactions, and ${\cal V}_e$=${\cal V}_i$.  It can be efficiently computed
using the ``lithium metal cohesive energy'' method widely used in the battery
theoretical literature, although most previous work neglects the possibility
of net surface charges, compensated by ions inside the liquid electrolyte,
which may affect parasitic reactions and Li$^+$ insertion kinetics.

But out-of-equilibrium conditions are also critical to many battery-related
phenomena such as SEI formation and electroplating.  Here ${\cal V}_e$, which
governs electron content at battery interfaces, is the correct definition of
voltage.  However, the work function-derived definition of ${\cal V}_e$ does
not fully account for spatial inhomogeneities.  Our solid state modeling
considerations form the basis for future AIMD simulations of thin film-coated
electrode/liquid electrolyte interfaces.

\section*{Supporting Information}
A supporting information (S.I.) document provides more details about the model
systems used, reports calculations on Li$_2$CO$_3$(001)/Au(111) interfaces
and  Li$_2$CO$_3$ bulk crystal decomposition thermodynamics considerations,
and gives a brief comparison between aqueous interfaces and battery electrolyte
interfaces.  This material is available free of charge via the Internet at
{\tt http://pubs.acs.org}.

\section*{Acknowledgement}

We thank Steve Harris, Sang Bok Lee, John Cumings, Yue Qi, Peter Feibelman,
and Ismaila Dabo
for useful discussions, and Nitin Kumar for performing prelminary CO$_3^{2-}$
breakdown simulations.  This work was supported by Nanostructures for
Electrical Energy Storage (NEES), an Energy Frontier Research Center funded by
the U.S.~Department of Energy, Office of Science, Office of Basic Energy
Sciences under Award Number DESC0001160.  Sandia National Laboratories is a
multiprogram laboratory managed and operated by Sandia Corporation, a wholly
owned subsidiary of Lockheed Martin Corporation, for the U.S.~Deparment of
Energy's National Nuclear Security Administration under contract
DE-AC04-94AL85000.

\bf{References}


\begin{references}
\bibitem{au_note}
See Ref.~\onlinecite{marzari}.
The coated Au(111) work functions are almost indistinguishale from
calculations where Au(111) slabs are symmetrically coated on both sides.
This verifies that no net charge resides on the bare Au(111) surface.
\bibitem{note7}
The spatially decomposed DOS plots Fig.~\ref{fig3},~\ref{fig5},~\ref{fig7},
and~\ref{fig9} follow Refs.~\onlinecite{voltage}.  They are constructed from
the VASP PROCAR output file which approximately decomposes $e^-$ spatial
distributions onto individual atoms.  Contributions above $\rho_c$=0.07
and $\rho_c$=0.05 are included for Li$_3$PO$_4$/Au and Li$_2$CO$_3$ plots.
Different criterion are used because $e^-$ are delocalized to different
extent at different interfaces; we stress all DOS decomposition schemes are
approximate.
\bibitem{delta_notes}
Eq.~\ref{delta} implies that a small, $-0.01$~$|e|$/\AA\, interfacial dipole
density is sufficient to reduce the voltage by 1.8~V.
\bibitem{pf6_notes}
Most monatomic anions like Cl$^-$ would lose an $e^-$ to the electrode
under UHV conditions.  Fortunately, the highest occupied molecular orbital
(HOMO) of PF$_6^-$ lies far below the Fermi level in all cases and permit
this strictly illustrative test.
\end{references}
\end{document}